\documentclass[aps,prl,
reprint,
amsmath,amssymb,
]{revtex4-1}

\usepackage{subfigure}
\usepackage{graphicx}
\usepackage{epstopdf}
\usepackage{dcolumn}
\usepackage{bm}
\usepackage{amsmath}
\usepackage{amssymb}

\preprint{APS/123-QED}

\begin{document}

\title{Self-Seeded Free-Electron Lasers with Orbital Angular Momentum}

\author{Jiawei Yan}   
\author{Gianluca Geloni}

\affiliation{%
European XFEL, Holzkoppel 4, 22869 Schenefeld, Germany
}%


\begin{abstract}
X-ray beams carrying orbital angular momentum (OAM) are an emerging tool for probing matter. Optical elements, such as spiral phase plates and zone plates, have been widely used to generate OAM light. However, these optics are challenging to use at x-ray free-electron lasers (XFELs) due to the high impinging intensities. Here, we propose a self-seeded FEL method to produce intense  x-ray vortices. Unlike passive filtering after amplification, an optical element will be used to introduce the helical phase in the linear regime, significantly reducing the thermal load on the optical element. The generated OAM pulse is then used as a seed  and significantly amplified. Theoretical analysis and numerical simulations demonstrate that the power of the OAM seed pulse can be amplified by more than two orders of magnitude, reaching peak powers of several tens of gigawatts. The proposed method paves the way for high-power and high-repetition-rate  OAM pulses of XFEL light.

\end{abstract}

\maketitle

Structured light could provide new perspectives on numerous physical phenomena. In particular, optical vortices carrying orbital angular momentum (OAM)~\cite{allen1992orbital}, characterized by a helical phase-front exp($il\phi$) where $\phi$ is the azimuthal coordinate and $l$ is the topological charge, are intensively studied. OAM light at visible and infrared wavelengths have been used in a wide range of fields~\cite{shen2019optical,he1995direct, grier2003revolution,sit2017high,ding2015quantum, furhapter2005spiral}. In the short wavelength regime, optical vortices are promising means to trigger new phenomena through light-matter interaction~\cite{picon2010transferring,de2020photoelectric}. X-ray beams with OAM have been proposed for research in quadrupolar x-ray dichroism experiments~\cite{van2007prediction}, photoionization experiments~\cite{picon2010photoionization}, resonant inelastic x-ray scattering~\cite{rury2013examining}, and magnetic helicoidal dichroism \cite{fanciulli2021electromagnetic,fanciulli2022observation}. Most recently, a new type of phase dichroism has been demonstrated to probe the real-space configuration of the antiferromagnetic ground state with x-ray beams carrying OAM~\cite{mccarter2022antiferromagnetic}. Moreover, the hard x-ray helical dichroism of disordered molecular media has been demonstrated~\cite{rouxel2022hard}. At present, x-ray OAM light is provided by synchrotron radiation facilities. The generation of high-intensity x-ray OAM beams by x-ray free-electron lasers (XFELs) could open new possibilities for studies requiring well-defined spatial intensity and phase variations. 

Modern XFELs deliver high-brightness pulses with durations spanning from tens of femtoseconds down to the attosecond range, enabling new research in a variety of scientific fields~\cite{Pellegrinireview, Huangreview}. The majority of  XFEL facilities worldwide~\cite{lcls,sacla,pal,decking2020mhz,swissFEL} employ the mechanism of self-amplified spontaneous emission (SASE)~\cite{sase}. The SASE process starts from the initial electron-beam shot noise, and allows operation over a wide spectral range, reaching sub-angstrom wavelengths. Self-seeding schemes~\cite{feldhaus1997possible, geloni2011novel} are used at several facilities to enhance the temporal coherence of FEL pulses, while increasing their spectral density. However, the transverse radiation profile of FELs operating at saturation is restricted to the fundamental FEL mode, which can usually be approximated by a Gaussian mode with no azimuthal phase-dependence.

The generation of OAM light at XFELs can be realized by using helical undulators or forming electron bunches with helical shapes. In the first case, the harmonic radiation in helical undulators has long been theoretically shown to carry a helical phase~\cite{colson1981nonlinear,geloni2007theory,hemsing2020coherent}, which was experimentally demonstrated only recently~\cite{bahrdt2013first,hemsing2014first,ribivc2017extreme}. However, the intensity of the harmonic emission is much weaker than that of the fundamental~\cite{geloni2007theory}. In order to deal with this issue, harmonic interaction between a seed laser and an electron beam in a helical undulator~\cite{hemsing2011generating,hemsing2013coherent} was further proposed to generate OAM radiation at the fundamental wavelength. In addition, schemes to generate OAM radiation by using an external seed laser with a proper transverse phase structure to form the electron bunches into a helical pattern~\cite{hemsing2012echo,ribivc2014generation} have been proposed. However, these methods are limited by the availability of external seed lasers, or by the harmonic conversion number, especially at the shortest wavelengths. Also, a transverse mode selection method using Bragg reflectors and longitudinal-transverse mode coupling~\cite{huang2021generating} has been proposed, but it requires an XFEL oscillator configuration. 

A more straightforward approach to generate x-ray OAM light is to employ optical elements, such as spiral phase plate (SPP)~\cite{peele2002observation,seiboth2019refractive}, spiral Fresnel zone plate (SZP)~\cite{sakdinawat2007soft,vila2014characterization}, and other diffractive optics~\cite{lee2019laguerre}, behind an undulator. In recent years, these optical elements have been well developed and have been experimentally demonstrated to produce soft and hard x-ray OAM light at synchrotron light sources. Recently, SPPs made from fused silica~\cite{seiboth2019refractive} have been successfully used to shape synchrotron radiation at photon energies of 8.2 keV, and are considered for potential use at XFELs. In addition, SZPs with silicon membranes have been successfully used at an FEL in the extreme-ultraviolet wavelength range with an efficiency of 17\% and a pulse energy of around 10 $\rm \mu J$~\cite{ribivc2017extreme}. However, for obtaining X-ray OAM radiation with high pulse energies, the efficiency and thermal loading of these optics can constitute a significant challenge, especially considering  the development of high-repetition-rate XFELs based on superconducting accelerators. 

In this letter we propose a relatively simple scheme to generate intense x-ray vortices that is free from the previous issues. The method is based on the widely used SASE mode of operation and does not require helical undulators or external seed laser systems, thus it can in principle be applied to all existing XFEL user facilities with minimal hardware addition, especially when in synergy with self-seeding setups. 

\begin{figure}[htp] 
	\centering 
	\includegraphics[width=1\linewidth]{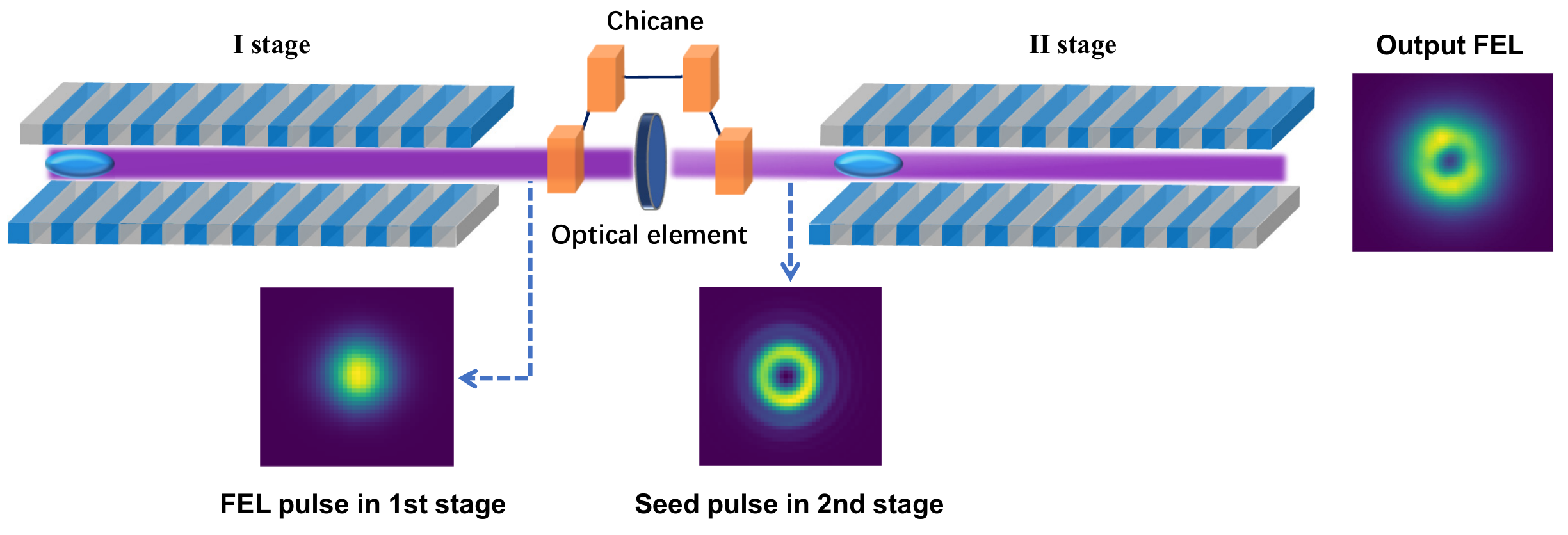}
	\caption{Schematic layout of the self-seeded FEL with OAM.}
	\label{figure1} 
\end{figure} 

The schematic layout of the proposed scheme referred to as self-seeded FEL with OAM (SSOAM), is shown in Fig.\ \ref{figure1}. Several undulator segments are employed as the first stage to generate a SASE FEL pulse in the linear regime. Then, an optical element, such as SPP or SZP, is used to imprint the helical phase of a low-order OAM mode onto the FEL beam, generating an OAM seed pulse. Finally, the OAM seed pulse is significantly amplified in the second stage undulator. It should be noted that high-order OAM modes with a topological charge larger than 2 cannot be well amplified due to stronger diffraction\cite{hemsing2008longitudinal,hemsing2012echo}. A small magnetic chicane located between the first and second stage is needed to detour the electron bunch, removes the microbunching introduced in the first stage, and results in a small delay, below ten femtoseconds, between the OAM seed and the electron bunch. Therefore, a relatively long electron bunch is requested in this scheme to ensure the amplification of the OAM seed in the second stage. In addition, fresh slice \cite{lutman2016fresh} or twin bunch operation \cite{marinelli2015high} can also be employed to obtain better performance. The drift section between the first stage and the optical element can be increased to make the transverse size of the FEL pulses hitting the optics larger; this further reduces the impinging intensity and therefore minimizes heat loading effects. In this case, it may be necessary to utilize an SZP or additional focusing optics before entering the second stage.

Previous theoretical analyses indicated that an OAM seed can be amplified by the FEL process \cite{hemsing2008longitudinal,hemsing2008virtual}. In the SSOAM scheme, however, the SASE pulse filtered by the optical element is not a pure OAM pulse. We can theoretically analyze how FEL modes evolve in this scheme by studying a three-step process. First, we expand the slowly varying envelope of the electric field in the frequency domain into a Fourier series in the azimuthal angle $\phi$ according to 
\begin{equation}
\widetilde{E}(z, r, \phi)=\sum_{n=-\infty}^{\infty} \widetilde{E}^{(n)}(z, r) \exp (-i n \phi),
\end{equation}
where z and r are the longitudinal and transverse coordinates with respect to the undulator axis. Each of the azimuthal harmonic is then decomposed into "self-reproducing" modes: 
\begin{align}
\widetilde{E}^{(n)}(\hat{z},\hat{r}) = \sum_j a_j^{(n)} \Phi_{nj}(\hat{r}) \exp(\lambda_j^{(n)} \hat{z})
\end{align}
with $\hat{r}=r/r_0$ and $\hat{z}=\Gamma z$. $\hat{r}$ and $\hat{z}$ are normalized versions of $z$ and $r$ according to the definition in \cite{saldin2000diffraction}, where $\Gamma$ is the gain parameter. Assuming no changes in the transverse profile and typical dimension $r_0$ of the electron beam, and the same gain parameter during the entire process, the normalization parameters will remain constant during all the steps, and the FEL eigenmodes obey the eigenvalue equation \cite{saldin2000diffraction}:
\begin{align}
\left[\Delta_n + g(\hat{r}, \lambda^{(n)}_j)\right]  \Phi_{nj}(\hat{r}) + 2i B  \lambda^{(n)}_j \Phi_{nj}(\hat{r}) = 0,
\end{align}
where we recognize that
\begin{align}
\Delta_n \equiv \left[\frac{\partial^2}{\partial \hat{r}^2}+\frac{1}{\hat{r}} \frac{\partial}{\partial \hat{r}} - \frac{n^2}{\hat{r}^2}\right]
\end{align}
is the Bessel differential operator. $\lambda^{(n)}_j$ is the eigenvalue. 
Here $g$ is a function of $\hat{r}$ and $\lambda^{(n)}_j$ defined in \cite{saldin2000diffraction} and $B=r_0^2 \Gamma \omega/c$.

During free-space propagation instead, the azimuthal harmonics of the electric field does not evolve in self-reproducing modes. Since

\begin{align}
\left[\nabla^2 + \frac{2 i \omega}{c} \frac{\partial}{\partial z}\right] \widetilde{E} = 0~,
\end{align}
after expansion in azimuthal harmonics we have
\begin{align}
\left[\Delta_n + 2 i B \frac{\partial}{\partial \hat{z}}\right]  \Phi_{nj}(\hat{r},\hat{z})  = 0.
\end{align}
In the first stage we have a first azimuthal series of FEL modes $\Phi^{(I)}_{nj}$ following:
\begin{align}
\left[\Delta_n + g(\hat{r}, \lambda^{(n)}_j)\right]  \Phi^{(I)}_{nj}(\hat{r}) + 2i B  \lambda^{(n)}_j \Phi^{(I)}_{nj}(\hat{r}) = 0.
\end{align}
The modes evolve in free space before filtering. We know the initial field, which is fixed by the modes $\Phi^{(I)}_{nj}$. Then, we call with  $\Phi^{(Ia)}_{nj}(\hat{z},\hat{r})$ the free-space propagated version of $\Phi^{(I)}_{nj}$, which obeys
\begin{align}
\left[\Delta_n + 2 i B \frac{\partial}{\partial \hat{z}}\right]  \Phi^{(Ia)}_{nj}(\hat{r},\hat{z})  = 0
\end{align}
For simplicity, here we assume that filtering happens directly at the exit of the first undulator. Therefore we neglect this first propagation step and filter directly $\Phi_{nj}^{(I)}$. The filtering amounts to a change in the phase, where a phase of exp($il\phi$) is introduced to the radiation. We can therefore write the filtered self-reproducing modes as  $\Phi^{(Ib)}_{n-l,j}(\hat{r}) = \Phi^{(I)}_{nj}$.
The modes $\Phi^{(Ib)}_{n-l,j}$ need now to be propagated in free space for a distance delta $\hat{z}$. We call with  $\Phi^{(Ic)}_{n-l,j}(\hat{z},\hat{r})$ the free-space propagated version of $\Phi^{(Ib)}_{n-l,j}$, which obeys

\begin{align}
\left[\Delta_{n-l} + 2 i B \frac{\partial}{\partial \hat{z}}\right]  \Phi^{(Ic)}_{n-l,j}(\hat{r},\hat{z})  = 0
\end{align}
Solving this equation for a propagation distance delta $\hat{z}$ we obtain



\begin{equation}
\begin{aligned}
\Phi_{n-l, j}^{(I c)}(\hat{r}, \delta \hat{z})=& \frac{i B}{2 \pi \delta \hat{z}} \int_{0}^{\infty} d \hat{r}^{\prime} \hat{r}^{\prime} \int_{0}^{2 \pi} d \delta \phi^{\prime} \\
&\exp \left[\frac { i B } { 2 \delta \hat { z } } \left(\hat{r}^{2}-2 \hat{r} \hat{r}^{\prime} \cos \left(\delta \phi^{\prime}\right)+\hat{r}^{\prime 2}\right)\right]\\
& \times \Phi_{n j}^{(I)}\left(\hat{r}^{\prime}\right) \exp \left[-i(n-l) \delta \phi^{\prime}\right]
\end{aligned}
\end{equation}


%
Performing first the integral over the phase we write
\begin{align}
\begin{split}
\Phi^{(Ic)}_{n-l,j}(\hat{r},\delta \hat{z})  = & \frac{i^{n-l+1} B}{ \delta \hat{z}} \exp\left(\frac{i B \hat{r}^2}{2  \delta \hat{z}}\right)\int_0^\infty d \hat{r}' \hat{r}' \\ &
\exp\left(\frac{i B \hat{r}'^2}{2  \delta \hat{z}}\right) \Phi^{(I)}_{nj}(\hat{r}') 
  J_{n-l}\left(-\frac{    \hat{r}\hat{r}'B}{  \delta \hat{z}}\right)
\end{split}
\end{align} 
%
%
%
%
%
%
%
Summarizing, the field at the entrance of the second stage amounts to
\begin{align}
\begin{split}
\widetilde{E}(\hat{r},\phi) = & \sum_{n=-\infty}^{\infty} \sum_j a_j^{(n)}  \exp(\lambda_j^{(n)} \hat{z}_1)\\& \Phi^{(Ic)}_{n-l,j}(\hat{r}) \exp[-i(n-l) \phi] 
\end{split}
\end{align}
One may approximate $j=1$ as fixed because we assume good mode selection in the first stage. Then, the contribution to the field at the entrance of the second stage can be approximated to

\begin{align}
\begin{split}
\widetilde{E}(\hat{r},\phi) = & \sum_{n=-\infty}^{\infty} a_1^{(n)} \exp(\lambda_1^{(n)} \hat{z}_1) \Phi^{(Ic)}_{n-l,1}(\hat{r}) \exp[-i (n-l) \phi] \\
 \equiv & \sum_{n=-\infty}^{\infty} A^{(n)} \Phi^{(Ic)}_{n-l,1}(\hat{r}) \exp[-i (n-l) \phi] 
\end{split}
\end{align}
The contribution for $n=0$ corresponds to the fundamental FEL mode and will be dominant. All the contributions for all values of $n$ will enter the second stage with a topological charge $h = -(n-l)$.

\begin{figure}[!htb]
	\centering
	\subfigure[]{\includegraphics*[width=0.47\linewidth]{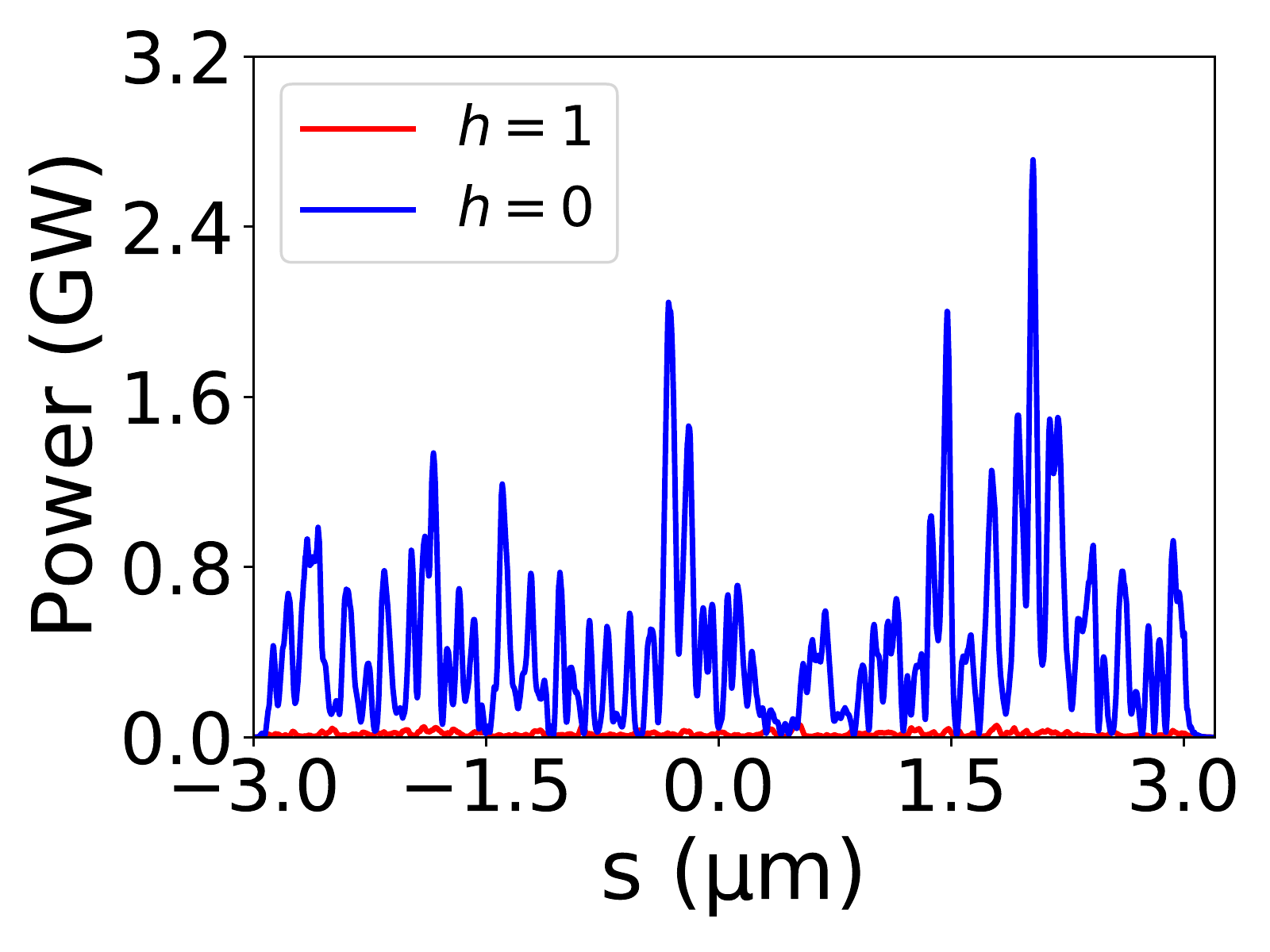}}
	\subfigure[]{\includegraphics*[width=0.47\linewidth]{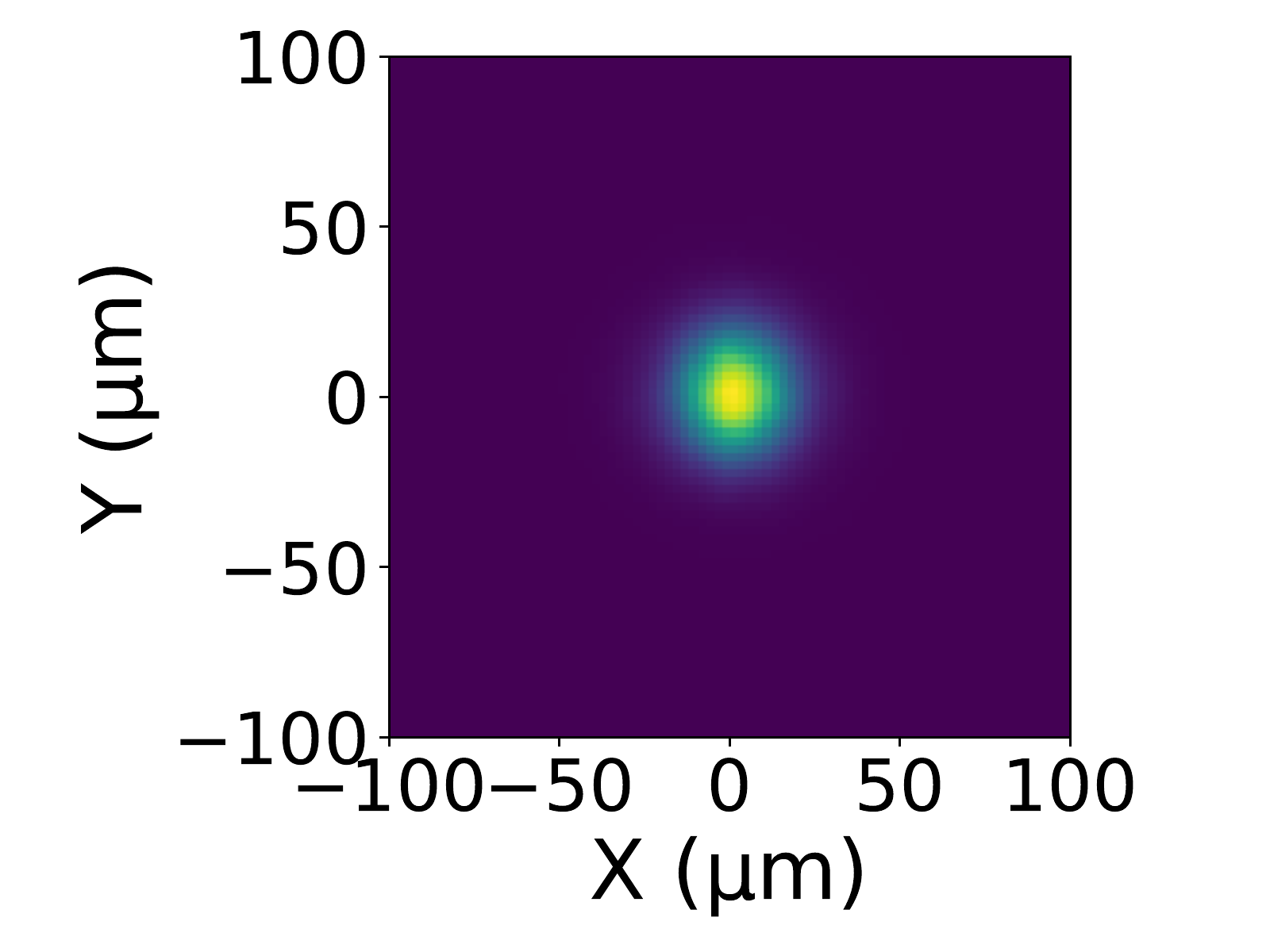}}
	\subfigure[]{\includegraphics*[width=0.47\linewidth]{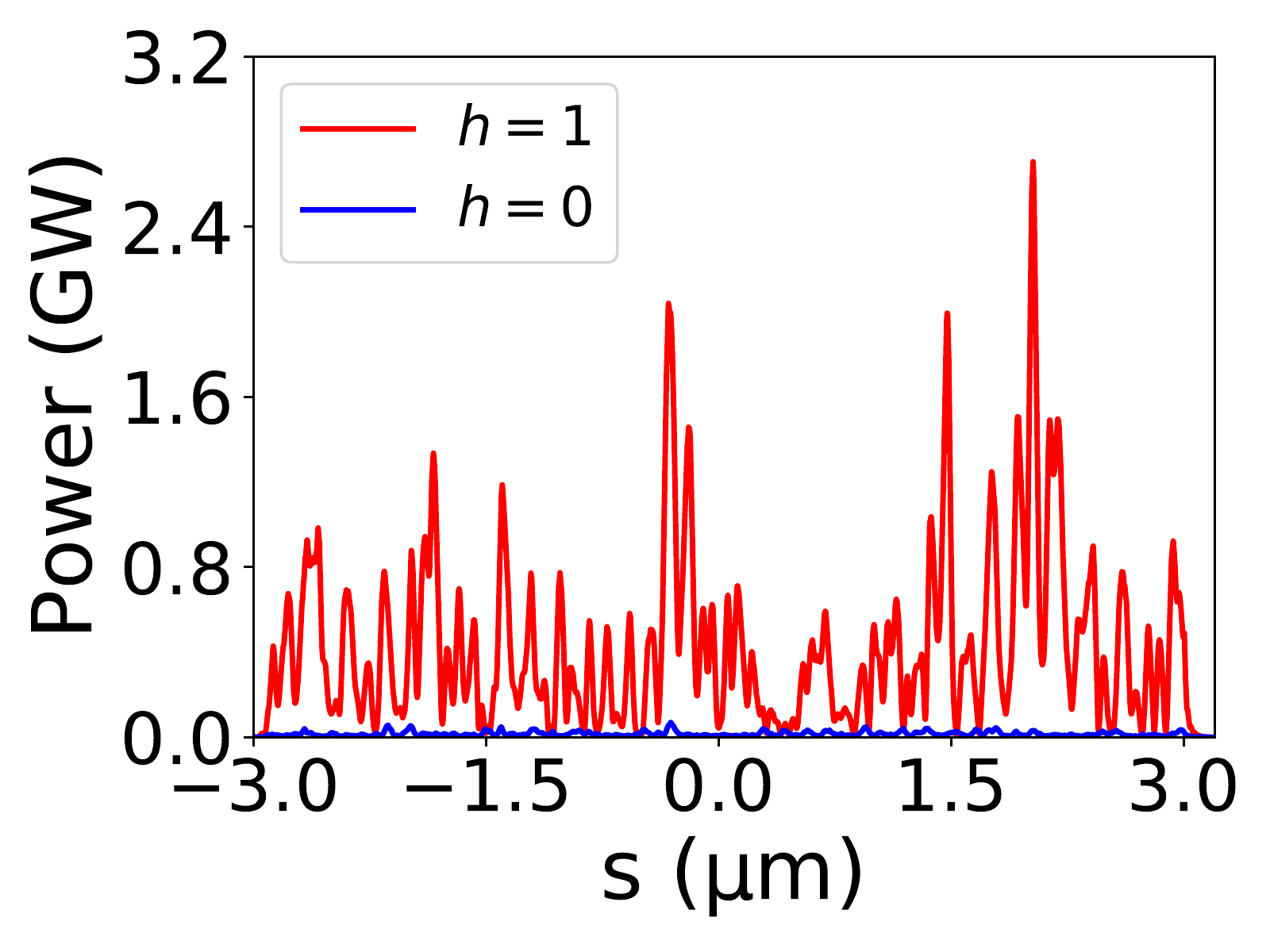}}
	\subfigure[]{\includegraphics*[width=0.47\linewidth]{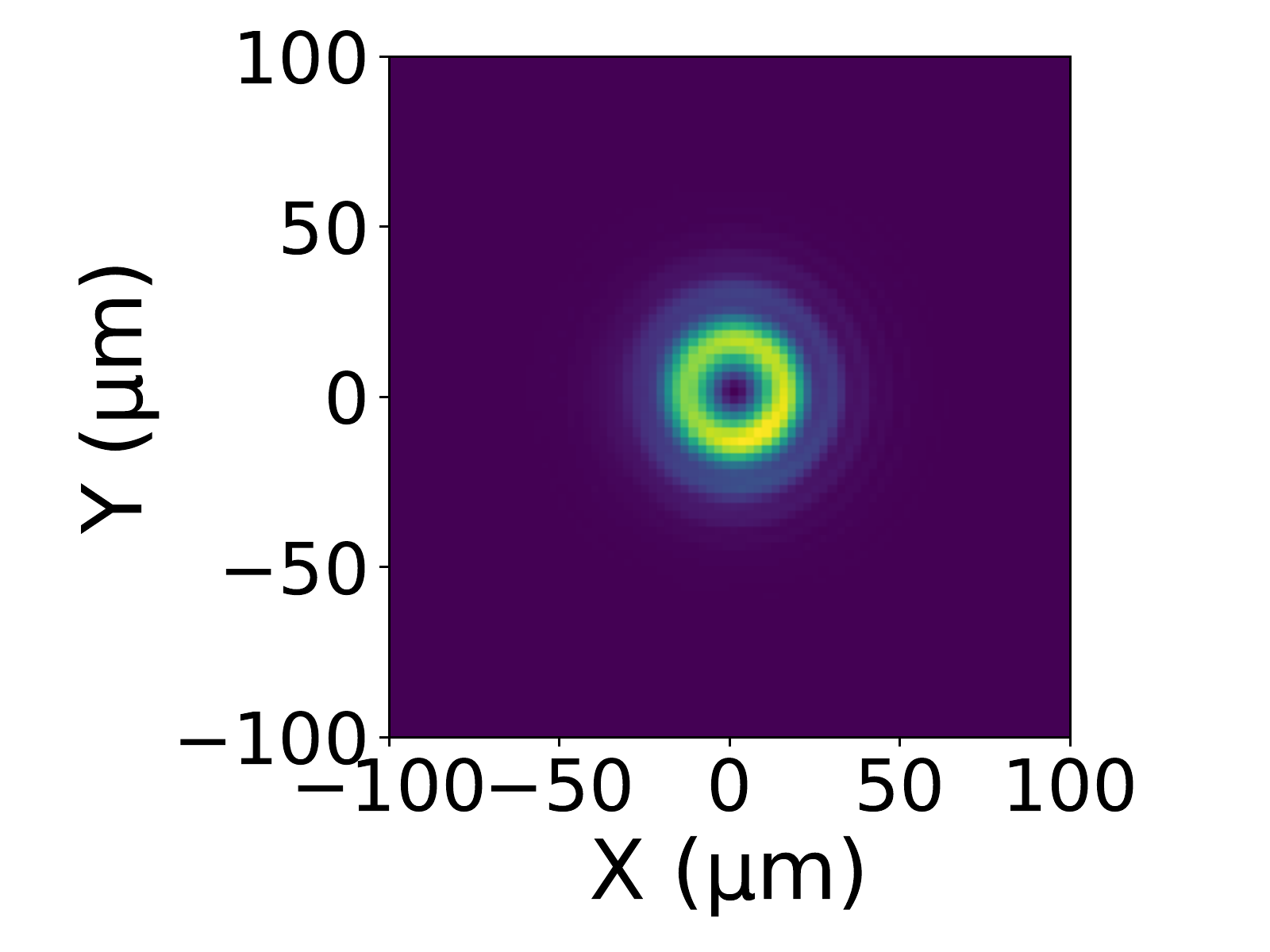}}
	\caption{Temporal power of the FEL pulse at (a) the end of the first stage and (c) the entrance of the second stage decomposed into OAM modes. The corresponding transverse profiles of the pulse at these two positions are shown in (b) and (d), respectively.}
	\label{stage2}
\end{figure}

Now, the FEL eigenmodes in the second stage obey

\begin{align}
\left[\Delta_{m} + g(\hat{r}, \lambda^{(II,m)}_k)\right]  \Phi^{(II)}_{m,k}(\hat{r}) + 2i B  \lambda^{(II,m)}_k  \Phi^{(II)}_{m,k}(\hat{r})= 0
\end{align}
and we can decompose

\begin{align}
\Phi^{(Ic)}_{n-l,1}(\hat{r},\delta \hat{z}) = \sum_k \alpha_k \Phi^{(II)}_{n-l,k} 
\end{align}
Since the orbital momentum should be conserved, each of the components $\Phi^{(II)}_{n-l,k}$ carries a fraction of the total OAM with  topological charge  $-(n-l)$. Each mode will be amplified at different rates for different values of $k$. Assuming good mode selection, the mode for $k=1$ will be dominant for the topological charge $-(n-l)$. However, all the FEL modes with fixed $-(n-l)$ and different $k$ will contribute to an OAM field distribution with the same topological charge given by 

\begin{align}
\begin{split}
\widetilde{E}^{(n-l)}(\hat{r},\phi) = &  A^{(n)}\sum_k \alpha_k \Phi^{(II)}_{n-l,k}  \exp[-i (n-l) \phi]\\& \exp(\lambda^{(II,n-l)}_k \hat{z}_2)
\end{split}
\end{align}
In this sense, these modes will not be in competition. The competition will come from modes with different topological charges. The expected main contribution will be for $m=-(n-l)=0$ and $k=1$ that corresponds to the main FEL eigenmode of the second stage.

To further illustrate the proposed scheme, a detailed example based on the parameters of European XFEL~\cite{decking2020mhz} is studied with GENESIS ~\cite{reiche1999genesis} simulations. A 14 GeV electron beam with a normalized emittance of 0.5 mm mrad, a bunch length of 20 fs, and a current flat-top profile of 5000 A is assumed here to produce 9 keV FEL pulses. The electron beam is sent to the first stage and generates SASE pulses with energies of a few microjoules or less. Then, an SPP is used to impart a helical phase of exp($i\phi$), corresponding to $l=1$, on the SASE pulse. We assumed the intensity of the pulse unchanged by the SPP. The distance between SPP and both undulator sections is set to 2 m. For simplicity, the SASE pulse interacts with a fresh electron beam with an unchanged bunch length in the second stage. 
\begin{figure}[!htb]
	\centering
	\includegraphics[width=0.8\linewidth]{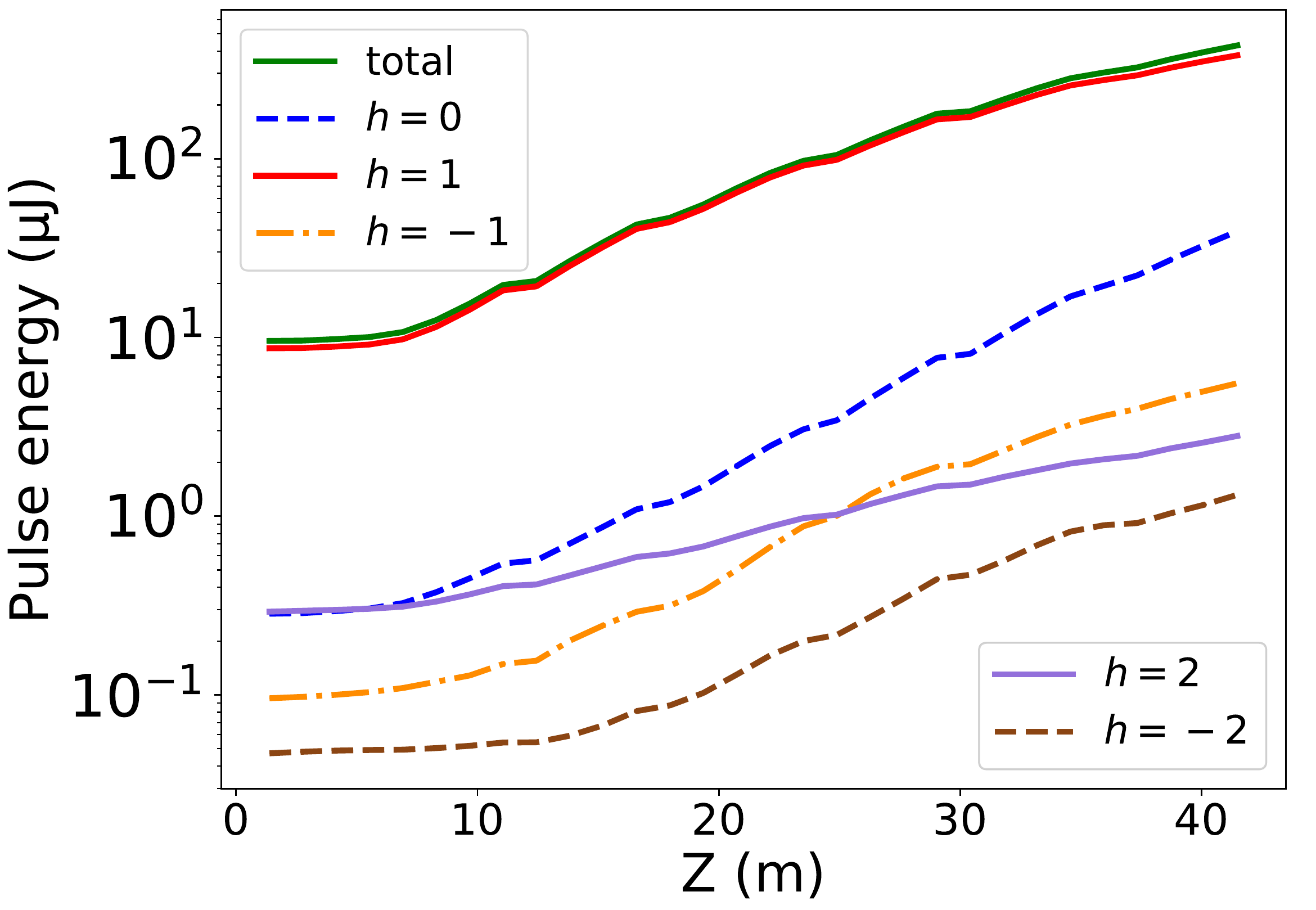}
	\caption{Gain curves of different OAM modes along the second stage. }
	\label{gaincurve}
\end{figure}

As indicated by the theoretical analysis, the $h = 0$ OAM mode of the SASE pulse is transferred to the $h = 1$ OAM mode while the the $h = -1$ mode is transferred to the $h = 0$ mode after the filtering. Therefore, the resonance and taper of the first stage undulator need to be optimized to achieve the largest possible ratio of the $h=0$ mode, while keeping the pulse energy small. Meanwhile, the undulator of the second stage needs to be optimized to achieve the strongest possible amplification of the $h = 1$ mode.

In the simulation, nine undulator segments with a length of 5 m and a $40$ mm period are employed as the first stage. The optimization of the first stage undulator suggests a slight reverse taper with $\Delta K/K =  -0.18\%$. In this case, a SASE pulse with a pulse energy of 9.58 $\rm \mu J$ is obtained at the end of the first stage. The mode decomposition of the pulse shows that the relative weight of the $h = 0$, $h = -1$, and $h = 1$ modes are 91\%, 3\%, and 3\%, respectively. The temporal power profile and transverse profile of the pulse are shown in Fig.~\ref{stage2} (a) and (b), respectively. The FEL pulse is then propagated sequentially through the free-space section, the SPP, and the second free-space section, before being amplified as a seed in the second stage. As shown in Fig.~\ref{stage2} (c) and (d), the seed pulse mainly consists of the $h = 1$ mode, with a ratio of 91\%. The ratio of the $h = 0$ mode of the seed pulse is 3\%.

\begin{figure}[!htb]
	\centering
	\subfigure[]{\includegraphics*[width=0.47\linewidth]{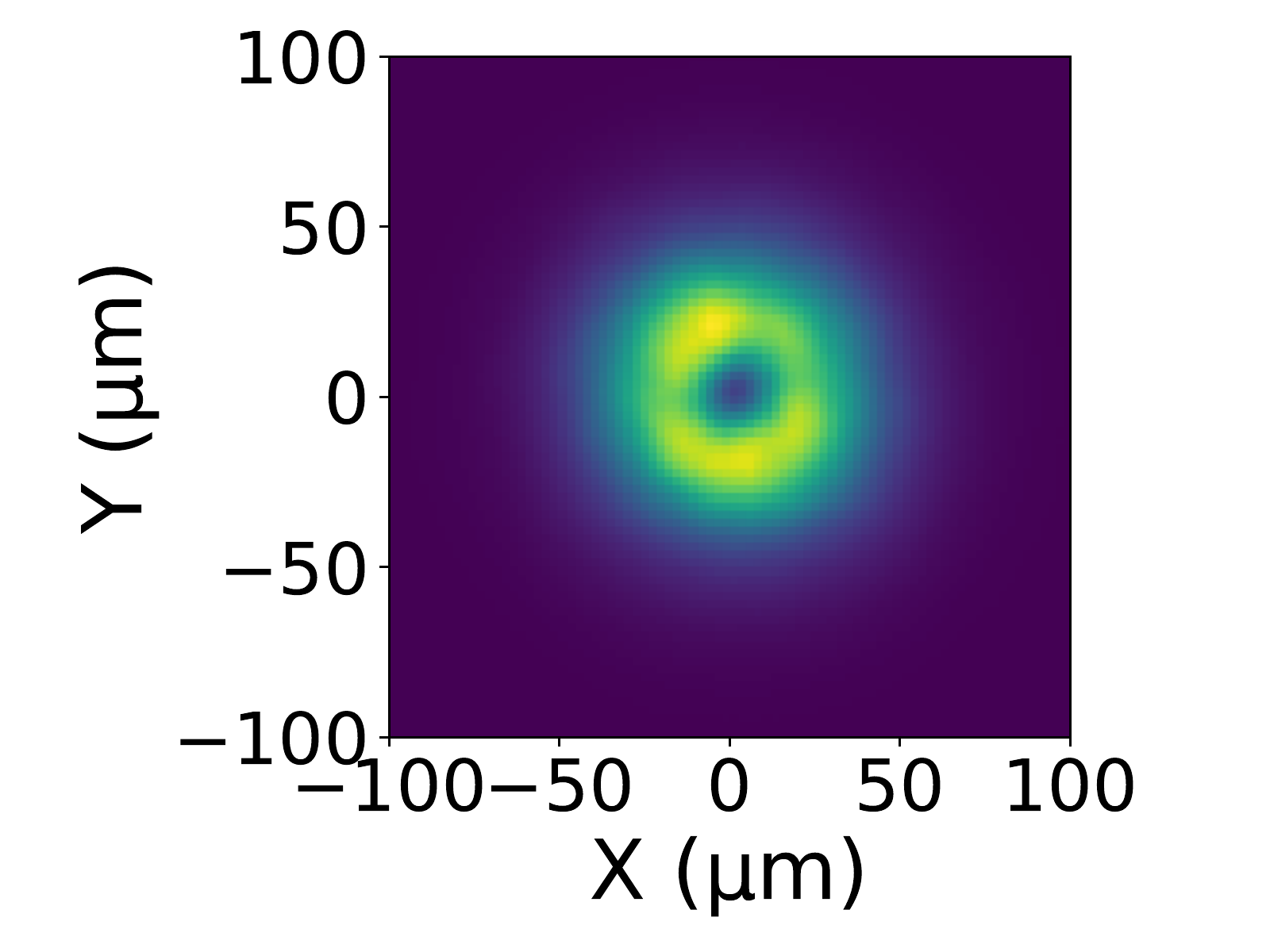}}
	\subfigure[]{\includegraphics*[width=0.47\linewidth]{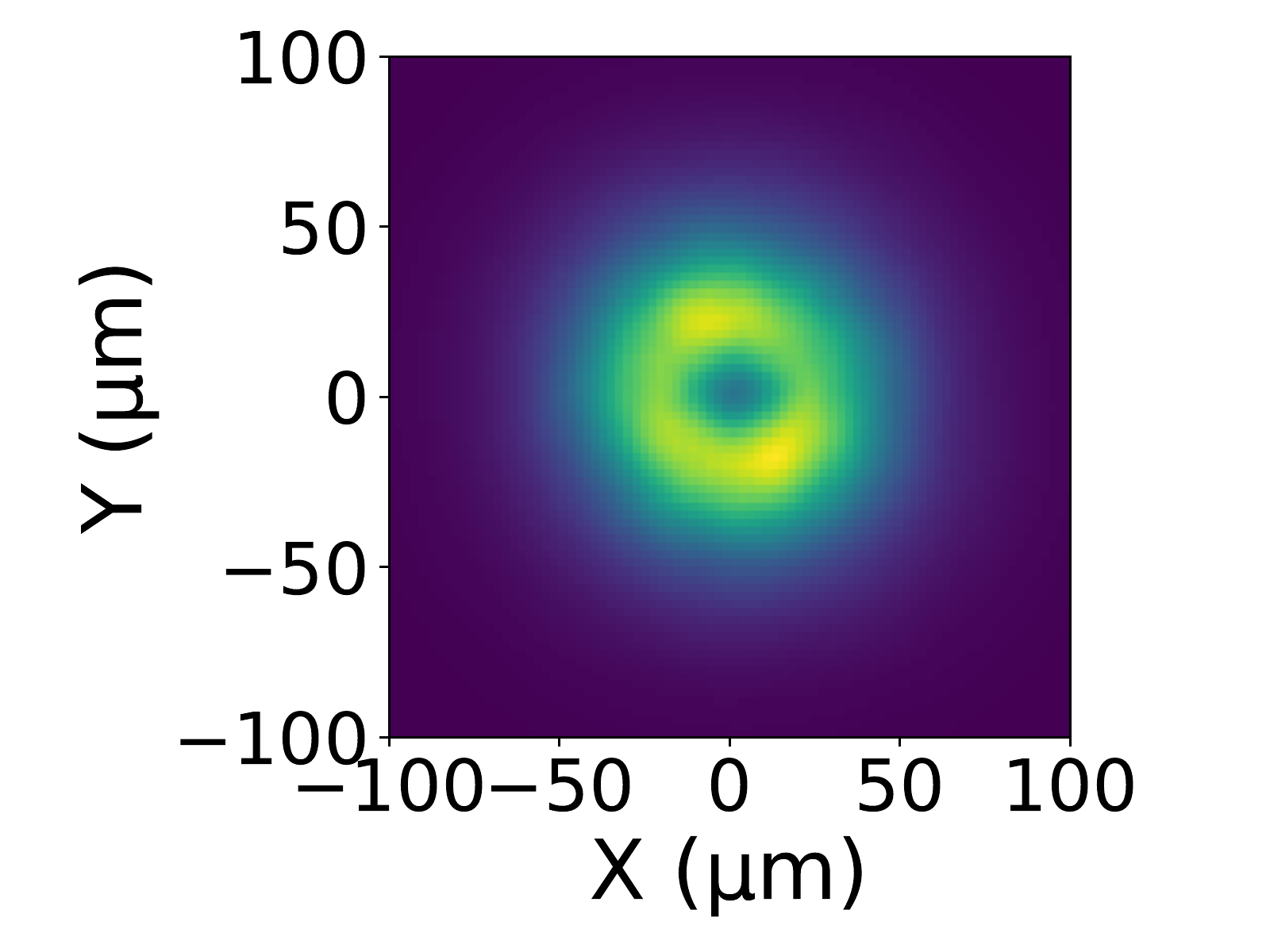}}
	\subfigure[]{\includegraphics*[width=0.47\linewidth]{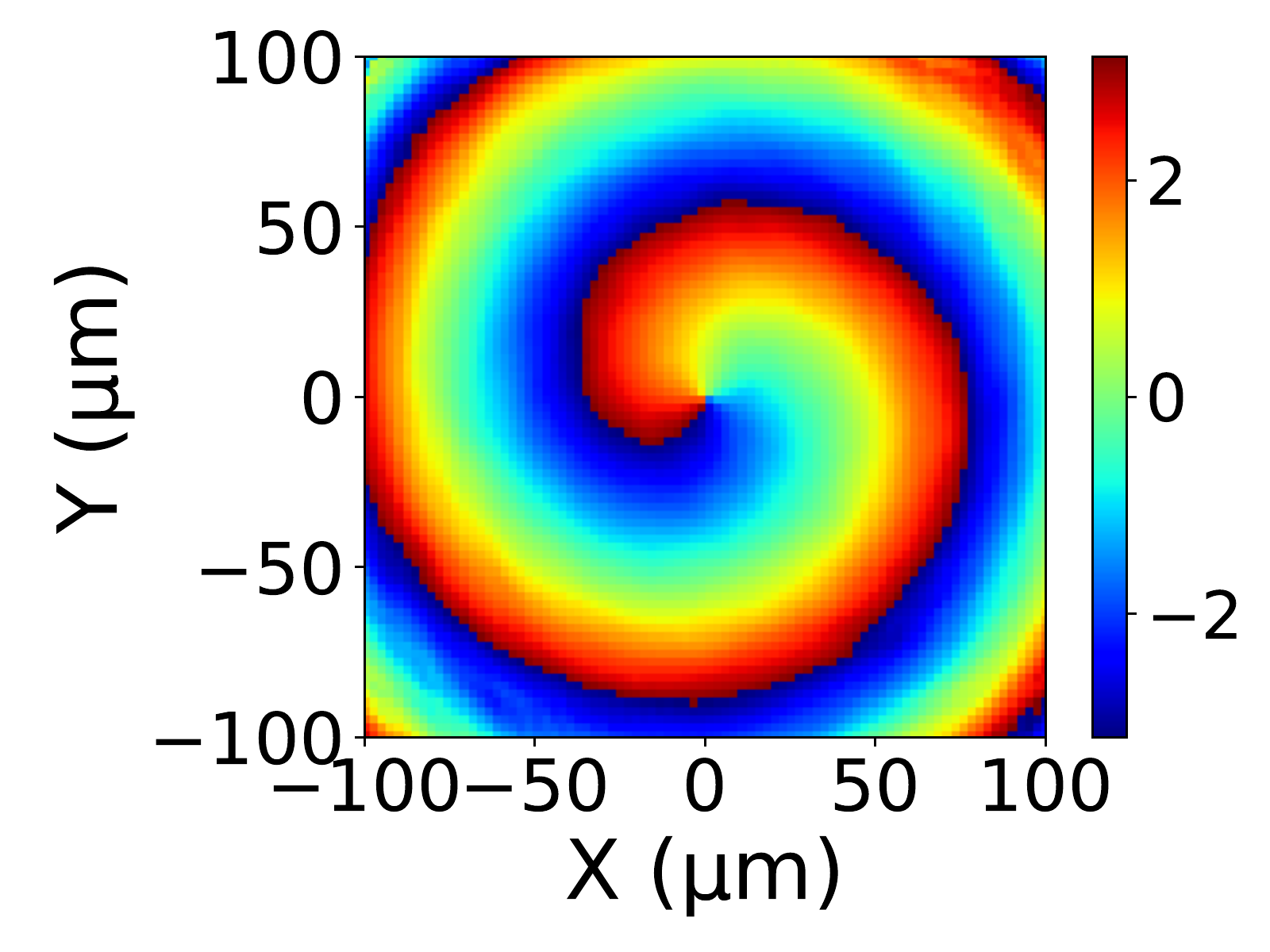}}
	\subfigure[]{\includegraphics*[width=0.47\linewidth]{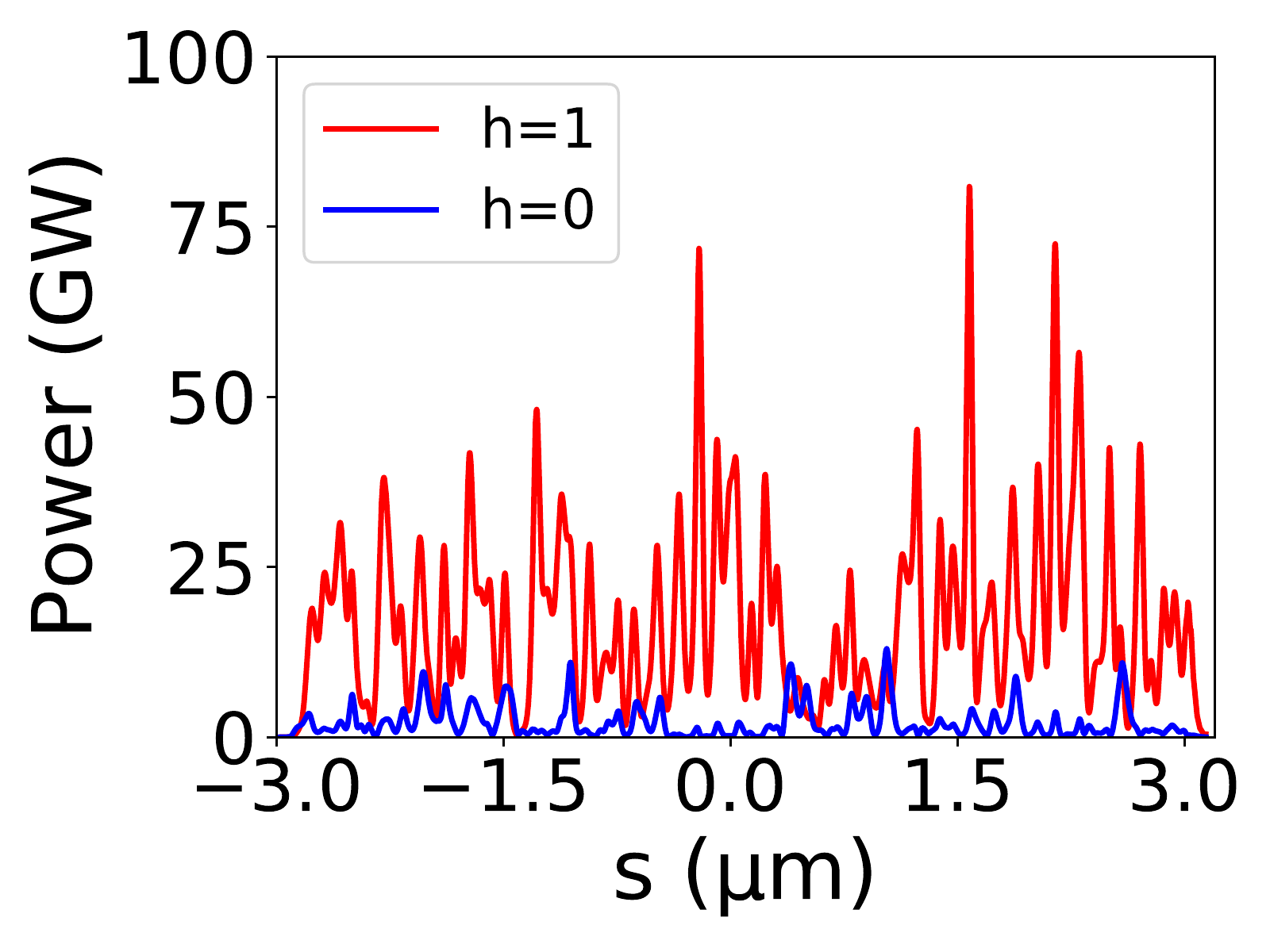}}
	\caption{Transverse profile of the FEL pulse at the end of (a) the fifth cell and (b) the second stage. (c) Transverse phase distribution at the position of maximum power in the pulse. (d) Temporal power profile of the pulse at the end of the second stage. }
	\label{tphase}
\end{figure}

The OAM seed pulse is amplified in the second stage, which consists of seven undulator segments. To ensure optimal gain for the $h = 1$ mode, the undulator taper is optimized to a strength of $\Delta K/K =  -0.06\%$. The gain curves of pulse and different OAM modes are shown in Fig.~\ref{gaincurve}. The pulse energy reaches 430.17 $\rm \mu J$ at the end of the undulator. The modes $h = 0$ and $h = 2$, which transform from $h = -1$ and $h = 1$ in the first stage respectively, have similar initial components in the second stage but grow at very different rates. At the exit of the second stage, the $h=1$ mode is amplified 43.5 fold and approaches saturation. The ratio of the $h = 1$ mode is first increased to 95\% and then slowly decreased to 88\%. The ratio of the $h = 0$ mode reaches 9\% at the end of the second stage. Fig.~\ref{tphase} (a) and (b) show the transverse profiles of the pulse at the exit of the fifth undulator cell, where the pulse energy is 184.44 $\rm \mu J$ and the ratio of the $h = 1$ mode is 93\%, and the end of the seventh cell, respectively. Fig.~\ref{tphase} (c) presents the transverse phase distribution at the position in the pulse with a peak power of 92 GW, located at 1.58 $\mu m$ in Fig.~\ref{tphase} (d).

\begin{figure}[!htb]
	\centering
	\subfigure[]{\includegraphics*[width=0.48\linewidth]{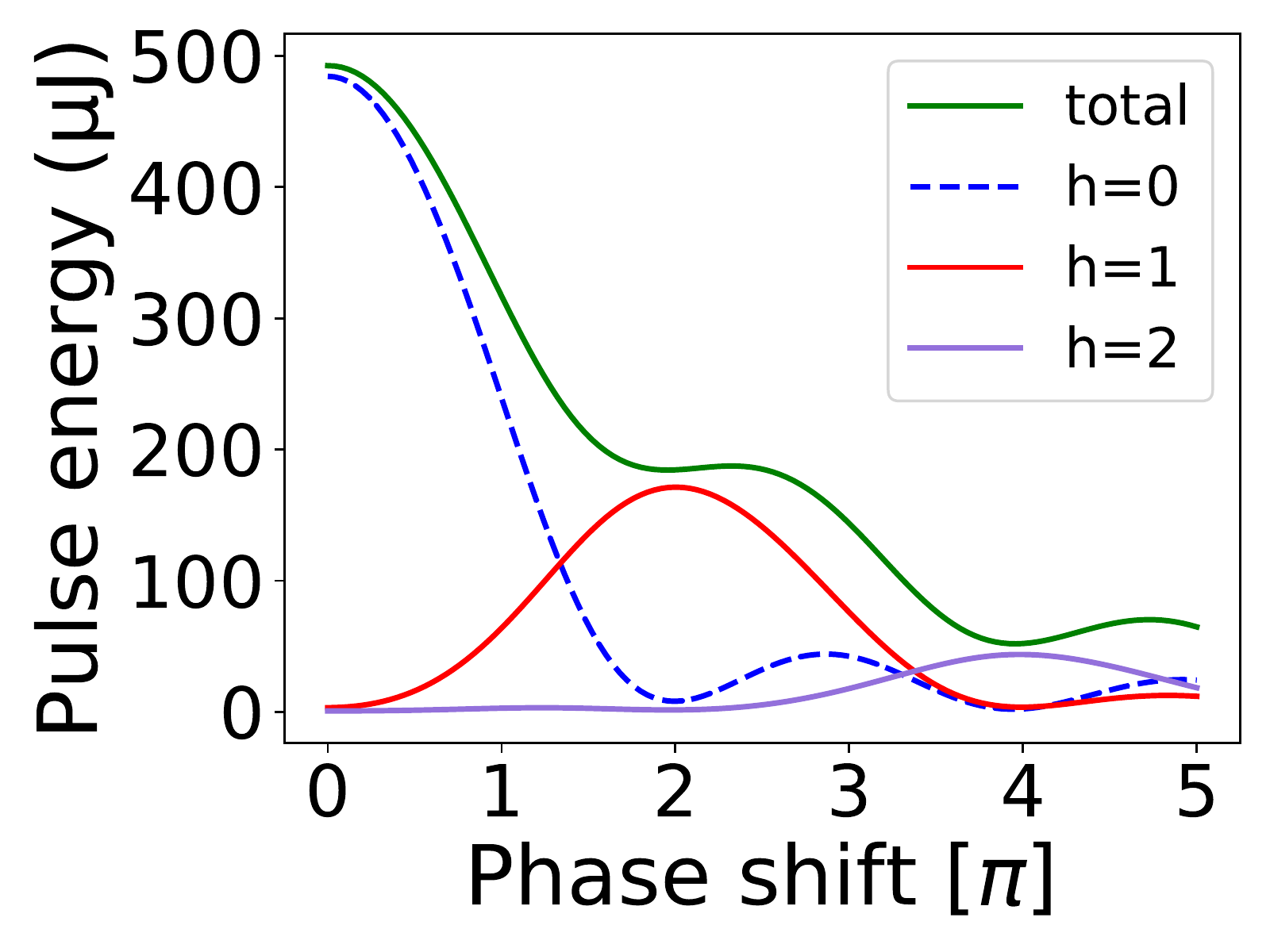}}
	\subfigure[]{\includegraphics*[width=0.48\linewidth]{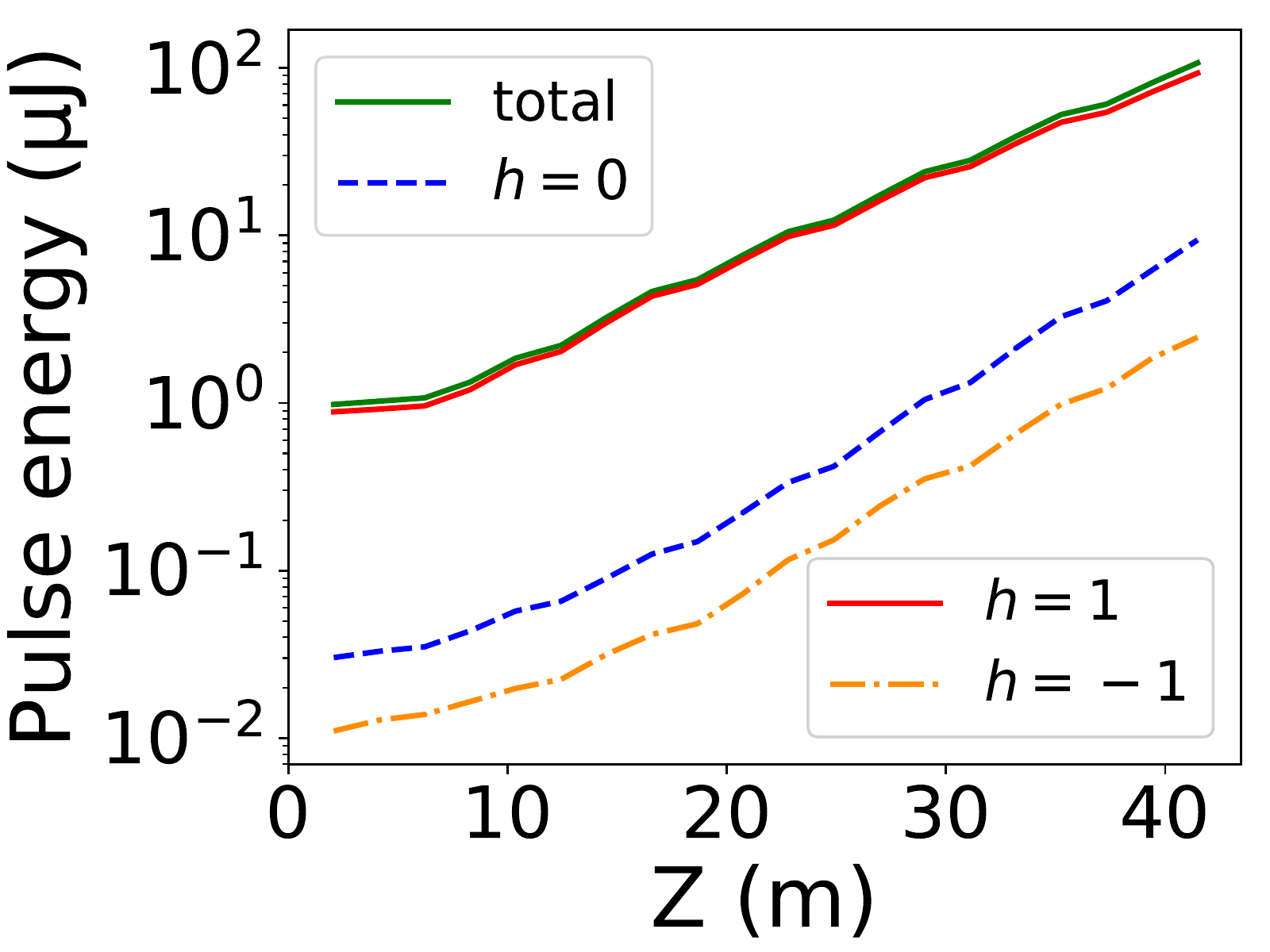}}
	\caption{(a) Variation of the different OAM modes with the change of the imprinted total phase shift. (b) Gain curves of different OAM modes at the second stage starting from a weaker seed pulse. }
	\label{weakcase}
\end{figure}

Fig.~\ref{weakcase} (a) presents the variation of the different OAM modes contained in the FEL pulse at the exit of the fifth undulator cell of the second stage as the total phase shift, i.e. $2 \pi l$, imprinted by the spiral phase plate changes. The results show that the proposed scheme can tolerate, to a certain extent, phase shift variations due to imperfections in optical element fabrication or changes in photon energy. With the introduction of phase shifts from $\rm 1.7 \pi$ to $\rm 2.4 \pi$, corresponding to $l=0.85$ to $l=1.2$, pulses with $h=1$ mode ratio exceeding 80\% can be obtained. FEL pulses with 84\% of the h=2 mode and pulse energy of 52 $\rm \mu J$ can be obtained when the introduced phase shift is $\rm 4 \pi$.

To simulate the effects of a weaker seed power, we assume that the use of an optical element still introduces exp($i\phi$) into the SASE pulse but with an efficiency of only 10\%, while the undulator settings remain unchanged. In comparison to Fig.~\ref{stage2} (c), the pulse energy of this seed pulse is reduced to 0.96 $\rm \mu J$. As shown in Fig.~\ref{weakcase} (b), the $h = 1$ mode can be amplified 107 fold in the second stage even if it starts with a much weaker seed pulse. At the exit of the seventh undulator cell, the pulse energy reaches 107.16 $\rm \mu J$ where the ratio of $h = 1$ and $h = 0$ modes is 87\% and 9\%, respectively.

In summary, we proposed a novel scheme to produce high-intensity short-wavelength vortices. Theoretical analysis and simulations demonstrate that the introduction of an OAM mode to an FEL beam using an optical element leads to changes in multiple transverse modes of the FEL beam, where the dominant OAM mode can be amplified more than two orders of magnitude. Since the SSOAM scheme is based on the SASE operation, it can be enabled in a wide range of wavelengths, including extreme ultraviolet, soft, and hard x-rays. In combination with the self-seeding scheme, introducing OAM to a weak and monochromatic seed pulse and further amplifying it, narrow bandwidth x-ray vortices can be generated. The proposed method paves the way for the realization of high-power and high-repetition-rate x-ray vortices and promises to open up new ways for probing matter with such light.


\bibliography{mybibfile}

\begin{thebibliography}{48}%
\makeatletter
\providecommand \@ifxundefined [1]{%
 \@ifx{#1\undefined}
}%
\providecommand \@ifnum [1]{%
 \ifnum #1\expandafter \@firstoftwo
 \else \expandafter \@secondoftwo
 \fi
}%
\providecommand \@ifx [1]{%
 \ifx #1\expandafter \@firstoftwo
 \else \expandafter \@secondoftwo
 \fi
}%
\providecommand \natexlab [1]{#1}%
\providecommand \enquote  [1]{``#1''}%
\providecommand \bibnamefont  [1]{#1}%
\providecommand \bibfnamefont [1]{#1}%
\providecommand \citenamefont [1]{#1}%
\providecommand \href@noop [0]{\@secondoftwo}%
\providecommand \href [0]{\begingroup \@sanitize@url \@href}%
\providecommand \@href[1]{\@@startlink{#1}\@@href}%
\providecommand \@@href[1]{\endgroup#1\@@endlink}%
\providecommand \@sanitize@url [0]{\catcode `\\12\catcode `\$12\catcode
  `\&12\catcode `\#12\catcode `\^12\catcode `\_12\catcode `\%12\relax}%
\providecommand \@@startlink[1]{}%
\providecommand \@@endlink[0]{}%
\providecommand \url  [0]{\begingroup\@sanitize@url \@url }%
\providecommand \@url [1]{\endgroup\@href {#1}{\urlprefix }}%
\providecommand \urlprefix  [0]{URL }%
\providecommand \Eprint [0]{\href }%
\providecommand \doibase [0]{http://dx.doi.org/}%
\providecommand \selectlanguage [0]{\@gobble}%
\providecommand \bibinfo  [0]{\@secondoftwo}%
\providecommand \bibfield  [0]{\@secondoftwo}%
\providecommand \translation [1]{[#1]}%
\providecommand \BibitemOpen [0]{}%
\providecommand \bibitemStop [0]{}%
\providecommand \bibitemNoStop [0]{.\EOS\space}%
\providecommand \EOS [0]{\spacefactor3000\relax}%
\providecommand \BibitemShut  [1]{\csname bibitem#1\endcsname}%
\let\auto@bib@innerbib\@empty
\bibitem [{\citenamefont {Allen}\ \emph {et~al.}(1992)\citenamefont {Allen},
  \citenamefont {Beijersbergen}, \citenamefont {Spreeuw},\ and\ \citenamefont
  {Woerdman}}]{allen1992orbital}%
  \BibitemOpen
  \bibfield  {author} {\bibinfo {author} {\bibfnamefont {L.}~\bibnamefont
  {Allen}}, \bibinfo {author} {\bibfnamefont {M.~W.}\ \bibnamefont
  {Beijersbergen}}, \bibinfo {author} {\bibfnamefont {R.}~\bibnamefont
  {Spreeuw}}, \ and\ \bibinfo {author} {\bibfnamefont {J.}~\bibnamefont
  {Woerdman}},\ }\href@noop {} {\bibfield  {journal} {\bibinfo  {journal}
  {Physical review A}\ }\textbf {\bibinfo {volume} {45}},\ \bibinfo {pages}
  {8185} (\bibinfo {year} {1992})}\BibitemShut {NoStop}%
\bibitem [{\citenamefont {Shen}\ \emph {et~al.}(2019)\citenamefont {Shen},
  \citenamefont {Wang}, \citenamefont {Xie}, \citenamefont {Min}, \citenamefont
  {Fu}, \citenamefont {Liu}, \citenamefont {Gong},\ and\ \citenamefont
  {Yuan}}]{shen2019optical}%
  \BibitemOpen
  \bibfield  {author} {\bibinfo {author} {\bibfnamefont {Y.}~\bibnamefont
  {Shen}}, \bibinfo {author} {\bibfnamefont {X.}~\bibnamefont {Wang}}, \bibinfo
  {author} {\bibfnamefont {Z.}~\bibnamefont {Xie}}, \bibinfo {author}
  {\bibfnamefont {C.}~\bibnamefont {Min}}, \bibinfo {author} {\bibfnamefont
  {X.}~\bibnamefont {Fu}}, \bibinfo {author} {\bibfnamefont {Q.}~\bibnamefont
  {Liu}}, \bibinfo {author} {\bibfnamefont {M.}~\bibnamefont {Gong}}, \ and\
  \bibinfo {author} {\bibfnamefont {X.}~\bibnamefont {Yuan}},\ }\href@noop {}
  {\bibfield  {journal} {\bibinfo  {journal} {Light: Science \& Applications}\
  }\textbf {\bibinfo {volume} {8}},\ \bibinfo {pages} {1} (\bibinfo {year}
  {2019})}\BibitemShut {NoStop}%
\bibitem [{\citenamefont {He}\ \emph {et~al.}(1995)\citenamefont {He},
  \citenamefont {Friese}, \citenamefont {Heckenberg},\ and\ \citenamefont
  {Rubinsztein-Dunlop}}]{he1995direct}%
  \BibitemOpen
  \bibfield  {author} {\bibinfo {author} {\bibfnamefont {H.}~\bibnamefont
  {He}}, \bibinfo {author} {\bibfnamefont {M.}~\bibnamefont {Friese}}, \bibinfo
  {author} {\bibfnamefont {N.}~\bibnamefont {Heckenberg}}, \ and\ \bibinfo
  {author} {\bibfnamefont {H.}~\bibnamefont {Rubinsztein-Dunlop}},\ }\href@noop
  {} {\bibfield  {journal} {\bibinfo  {journal} {Physical Review Letters}\
  }\textbf {\bibinfo {volume} {75}},\ \bibinfo {pages} {826} (\bibinfo {year}
  {1995})}\BibitemShut {NoStop}%
\bibitem [{\citenamefont {Grier}(2003)}]{grier2003revolution}%
  \BibitemOpen
  \bibfield  {author} {\bibinfo {author} {\bibfnamefont {D.~G.}\ \bibnamefont
  {Grier}},\ }\href@noop {} {\bibfield  {journal} {\bibinfo  {journal}
  {Nature}\ }\textbf {\bibinfo {volume} {424}},\ \bibinfo {pages} {810}
  (\bibinfo {year} {2003})}\BibitemShut {NoStop}%
\bibitem [{\citenamefont {Sit}\ \emph {et~al.}(2017)\citenamefont {Sit},
  \citenamefont {Bouchard}, \citenamefont {Fickler}, \citenamefont
  {Gagnon-Bischoff}, \citenamefont {Larocque}, \citenamefont {Heshami},
  \citenamefont {Elser}, \citenamefont {Peuntinger}, \citenamefont
  {G{\"u}nthner}, \citenamefont {Heim} \emph {et~al.}}]{sit2017high}%
  \BibitemOpen
  \bibfield  {author} {\bibinfo {author} {\bibfnamefont {A.}~\bibnamefont
  {Sit}}, \bibinfo {author} {\bibfnamefont {F.}~\bibnamefont {Bouchard}},
  \bibinfo {author} {\bibfnamefont {R.}~\bibnamefont {Fickler}}, \bibinfo
  {author} {\bibfnamefont {J.}~\bibnamefont {Gagnon-Bischoff}}, \bibinfo
  {author} {\bibfnamefont {H.}~\bibnamefont {Larocque}}, \bibinfo {author}
  {\bibfnamefont {K.}~\bibnamefont {Heshami}}, \bibinfo {author} {\bibfnamefont
  {D.}~\bibnamefont {Elser}}, \bibinfo {author} {\bibfnamefont
  {C.}~\bibnamefont {Peuntinger}}, \bibinfo {author} {\bibfnamefont
  {K.}~\bibnamefont {G{\"u}nthner}}, \bibinfo {author} {\bibfnamefont
  {B.}~\bibnamefont {Heim}},  \emph {et~al.},\ }\href@noop {} {\bibfield
  {journal} {\bibinfo  {journal} {Optica}\ }\textbf {\bibinfo {volume} {4}},\
  \bibinfo {pages} {1006} (\bibinfo {year} {2017})}\BibitemShut {NoStop}%
\bibitem [{\citenamefont {Ding}\ \emph {et~al.}(2015)\citenamefont {Ding},
  \citenamefont {Zhang}, \citenamefont {Zhou}, \citenamefont {Shi},
  \citenamefont {Xiang}, \citenamefont {Wang}, \citenamefont {Jiang},
  \citenamefont {Shi},\ and\ \citenamefont {Guo}}]{ding2015quantum}%
  \BibitemOpen
  \bibfield  {author} {\bibinfo {author} {\bibfnamefont {D.-S.}\ \bibnamefont
  {Ding}}, \bibinfo {author} {\bibfnamefont {W.}~\bibnamefont {Zhang}},
  \bibinfo {author} {\bibfnamefont {Z.-Y.}\ \bibnamefont {Zhou}}, \bibinfo
  {author} {\bibfnamefont {S.}~\bibnamefont {Shi}}, \bibinfo {author}
  {\bibfnamefont {G.-Y.}\ \bibnamefont {Xiang}}, \bibinfo {author}
  {\bibfnamefont {X.-S.}\ \bibnamefont {Wang}}, \bibinfo {author}
  {\bibfnamefont {Y.-K.}\ \bibnamefont {Jiang}}, \bibinfo {author}
  {\bibfnamefont {B.-S.}\ \bibnamefont {Shi}}, \ and\ \bibinfo {author}
  {\bibfnamefont {G.-C.}\ \bibnamefont {Guo}},\ }\href@noop {} {\bibfield
  {journal} {\bibinfo  {journal} {Physical Review Letters}\ }\textbf {\bibinfo
  {volume} {114}},\ \bibinfo {pages} {050502} (\bibinfo {year}
  {2015})}\BibitemShut {NoStop}%
\bibitem [{\citenamefont {F{\"u}rhapter}\ \emph {et~al.}(2005)\citenamefont
  {F{\"u}rhapter}, \citenamefont {Jesacher}, \citenamefont {Bernet},\ and\
  \citenamefont {Ritsch-Marte}}]{furhapter2005spiral}%
  \BibitemOpen
  \bibfield  {author} {\bibinfo {author} {\bibfnamefont {S.}~\bibnamefont
  {F{\"u}rhapter}}, \bibinfo {author} {\bibfnamefont {A.}~\bibnamefont
  {Jesacher}}, \bibinfo {author} {\bibfnamefont {S.}~\bibnamefont {Bernet}}, \
  and\ \bibinfo {author} {\bibfnamefont {M.}~\bibnamefont {Ritsch-Marte}},\
  }\href@noop {} {\bibfield  {journal} {\bibinfo  {journal} {Optics Express}\
  }\textbf {\bibinfo {volume} {13}},\ \bibinfo {pages} {689} (\bibinfo {year}
  {2005})}\BibitemShut {NoStop}%
\bibitem [{\citenamefont {Pic{\'o}n}\ \emph
  {et~al.}(2010{\natexlab{a}})\citenamefont {Pic{\'o}n}, \citenamefont
  {Benseny}, \citenamefont {Mompart}, \citenamefont {de~Aldana}, \citenamefont
  {Plaja}, \citenamefont {Calvo},\ and\ \citenamefont
  {Roso}}]{picon2010transferring}%
  \BibitemOpen
  \bibfield  {author} {\bibinfo {author} {\bibfnamefont {A.}~\bibnamefont
  {Pic{\'o}n}}, \bibinfo {author} {\bibfnamefont {A.}~\bibnamefont {Benseny}},
  \bibinfo {author} {\bibfnamefont {J.}~\bibnamefont {Mompart}}, \bibinfo
  {author} {\bibfnamefont {J.~V.}\ \bibnamefont {de~Aldana}}, \bibinfo {author}
  {\bibfnamefont {L.}~\bibnamefont {Plaja}}, \bibinfo {author} {\bibfnamefont
  {G.~F.}\ \bibnamefont {Calvo}}, \ and\ \bibinfo {author} {\bibfnamefont
  {L.}~\bibnamefont {Roso}},\ }\href@noop {} {\bibfield  {journal} {\bibinfo
  {journal} {New Journal of Physics}\ }\textbf {\bibinfo {volume} {12}},\
  \bibinfo {pages} {083053} (\bibinfo {year} {2010}{\natexlab{a}})}\BibitemShut
  {NoStop}%
\bibitem [{\citenamefont {De~Ninno}\ \emph {et~al.}(2020)\citenamefont
  {De~Ninno}, \citenamefont {W{\"a}tzel}, \citenamefont {Ribi{\v{c}}},
  \citenamefont {Allaria}, \citenamefont {Coreno}, \citenamefont {Danailov},
  \citenamefont {David}, \citenamefont {Demidovich}, \citenamefont {Di~Fraia},
  \citenamefont {Giannessi} \emph {et~al.}}]{de2020photoelectric}%
  \BibitemOpen
  \bibfield  {author} {\bibinfo {author} {\bibfnamefont {G.}~\bibnamefont
  {De~Ninno}}, \bibinfo {author} {\bibfnamefont {J.}~\bibnamefont
  {W{\"a}tzel}}, \bibinfo {author} {\bibfnamefont {P.~R.}\ \bibnamefont
  {Ribi{\v{c}}}}, \bibinfo {author} {\bibfnamefont {E.}~\bibnamefont
  {Allaria}}, \bibinfo {author} {\bibfnamefont {M.}~\bibnamefont {Coreno}},
  \bibinfo {author} {\bibfnamefont {M.~B.}\ \bibnamefont {Danailov}}, \bibinfo
  {author} {\bibfnamefont {C.}~\bibnamefont {David}}, \bibinfo {author}
  {\bibfnamefont {A.}~\bibnamefont {Demidovich}}, \bibinfo {author}
  {\bibfnamefont {M.}~\bibnamefont {Di~Fraia}}, \bibinfo {author}
  {\bibfnamefont {L.}~\bibnamefont {Giannessi}},  \emph {et~al.},\ }\href@noop
  {} {\bibfield  {journal} {\bibinfo  {journal} {Nature Photonics}\ }\textbf
  {\bibinfo {volume} {14}},\ \bibinfo {pages} {554} (\bibinfo {year}
  {2020})}\BibitemShut {NoStop}%
\bibitem [{\citenamefont {van Veenendaal}\ and\ \citenamefont
  {McNulty}(2007)}]{van2007prediction}%
  \BibitemOpen
  \bibfield  {author} {\bibinfo {author} {\bibfnamefont {M.}~\bibnamefont {van
  Veenendaal}}\ and\ \bibinfo {author} {\bibfnamefont {I.}~\bibnamefont
  {McNulty}},\ }\href@noop {} {\bibfield  {journal} {\bibinfo  {journal}
  {Physical Review Letters}\ }\textbf {\bibinfo {volume} {98}},\ \bibinfo
  {pages} {157401} (\bibinfo {year} {2007})}\BibitemShut {NoStop}%
\bibitem [{\citenamefont {Pic{\'o}n}\ \emph
  {et~al.}(2010{\natexlab{b}})\citenamefont {Pic{\'o}n}, \citenamefont
  {Mompart}, \citenamefont {de~Aldana}, \citenamefont {Plaja}, \citenamefont
  {Calvo},\ and\ \citenamefont {Roso}}]{picon2010photoionization}%
  \BibitemOpen
  \bibfield  {author} {\bibinfo {author} {\bibfnamefont {A.}~\bibnamefont
  {Pic{\'o}n}}, \bibinfo {author} {\bibfnamefont {J.}~\bibnamefont {Mompart}},
  \bibinfo {author} {\bibfnamefont {J.~V.}\ \bibnamefont {de~Aldana}}, \bibinfo
  {author} {\bibfnamefont {L.}~\bibnamefont {Plaja}}, \bibinfo {author}
  {\bibfnamefont {G.}~\bibnamefont {Calvo}}, \ and\ \bibinfo {author}
  {\bibfnamefont {L.}~\bibnamefont {Roso}},\ }\href@noop {} {\bibfield
  {journal} {\bibinfo  {journal} {Optics Express}\ }\textbf {\bibinfo {volume}
  {18}},\ \bibinfo {pages} {3660} (\bibinfo {year}
  {2010}{\natexlab{b}})}\BibitemShut {NoStop}%
\bibitem [{\citenamefont {Rury}(2013)}]{rury2013examining}%
  \BibitemOpen
  \bibfield  {author} {\bibinfo {author} {\bibfnamefont {A.~S.}\ \bibnamefont
  {Rury}},\ }\href@noop {} {\bibfield  {journal} {\bibinfo  {journal} {Physical
  Review A}\ }\textbf {\bibinfo {volume} {87}},\ \bibinfo {pages} {043408}
  (\bibinfo {year} {2013})}\BibitemShut {NoStop}%
\bibitem [{\citenamefont {Fanciulli}\ \emph {et~al.}(2021)\citenamefont
  {Fanciulli}, \citenamefont {Bresteau}, \citenamefont {Vimal}, \citenamefont
  {Luttmann}, \citenamefont {Sacchi},\ and\ \citenamefont
  {Ruchon}}]{fanciulli2021electromagnetic}%
  \BibitemOpen
  \bibfield  {author} {\bibinfo {author} {\bibfnamefont {M.}~\bibnamefont
  {Fanciulli}}, \bibinfo {author} {\bibfnamefont {D.}~\bibnamefont {Bresteau}},
  \bibinfo {author} {\bibfnamefont {M.}~\bibnamefont {Vimal}}, \bibinfo
  {author} {\bibfnamefont {M.}~\bibnamefont {Luttmann}}, \bibinfo {author}
  {\bibfnamefont {M.}~\bibnamefont {Sacchi}}, \ and\ \bibinfo {author}
  {\bibfnamefont {T.}~\bibnamefont {Ruchon}},\ }\href@noop {} {\bibfield
  {journal} {\bibinfo  {journal} {Physical Review A}\ }\textbf {\bibinfo
  {volume} {103}},\ \bibinfo {pages} {013501} (\bibinfo {year}
  {2021})}\BibitemShut {NoStop}%
\bibitem [{\citenamefont {Fanciulli}\ \emph {et~al.}(2022)\citenamefont
  {Fanciulli}, \citenamefont {Pancaldi}, \citenamefont {Pedersoli},
  \citenamefont {Vimal}, \citenamefont {Bresteau}, \citenamefont {Luttmann},
  \citenamefont {De~Angelis}, \citenamefont {Ribi{\v{c}}}, \citenamefont
  {R{\"o}sner}, \citenamefont {David} \emph
  {et~al.}}]{fanciulli2022observation}%
  \BibitemOpen
  \bibfield  {author} {\bibinfo {author} {\bibfnamefont {M.}~\bibnamefont
  {Fanciulli}}, \bibinfo {author} {\bibfnamefont {M.}~\bibnamefont {Pancaldi}},
  \bibinfo {author} {\bibfnamefont {E.}~\bibnamefont {Pedersoli}}, \bibinfo
  {author} {\bibfnamefont {M.}~\bibnamefont {Vimal}}, \bibinfo {author}
  {\bibfnamefont {D.}~\bibnamefont {Bresteau}}, \bibinfo {author}
  {\bibfnamefont {M.}~\bibnamefont {Luttmann}}, \bibinfo {author}
  {\bibfnamefont {D.}~\bibnamefont {De~Angelis}}, \bibinfo {author}
  {\bibfnamefont {P.~R.}\ \bibnamefont {Ribi{\v{c}}}}, \bibinfo {author}
  {\bibfnamefont {B.}~\bibnamefont {R{\"o}sner}}, \bibinfo {author}
  {\bibfnamefont {C.}~\bibnamefont {David}},  \emph {et~al.},\ }\href@noop {}
  {\bibfield  {journal} {\bibinfo  {journal} {Physical Review Letters}\
  }\textbf {\bibinfo {volume} {128}},\ \bibinfo {pages} {077401} (\bibinfo
  {year} {2022})}\BibitemShut {NoStop}%
\bibitem [{\citenamefont {McCarter}\ \emph {et~al.}(2022)\citenamefont
  {McCarter}, \citenamefont {Saleheen}, \citenamefont {Singh}, \citenamefont
  {Tumbleson}, \citenamefont {Woods}, \citenamefont {Tremsin}, \citenamefont
  {Scholl}, \citenamefont {De~Long}, \citenamefont {Hastings}, \citenamefont
  {Morley} \emph {et~al.}}]{mccarter2022antiferromagnetic}%
  \BibitemOpen
  \bibfield  {author} {\bibinfo {author} {\bibfnamefont {M.~R.}\ \bibnamefont
  {McCarter}}, \bibinfo {author} {\bibfnamefont {A.~I.}\ \bibnamefont
  {Saleheen}}, \bibinfo {author} {\bibfnamefont {A.}~\bibnamefont {Singh}},
  \bibinfo {author} {\bibfnamefont {R.}~\bibnamefont {Tumbleson}}, \bibinfo
  {author} {\bibfnamefont {J.~S.}\ \bibnamefont {Woods}}, \bibinfo {author}
  {\bibfnamefont {A.~S.}\ \bibnamefont {Tremsin}}, \bibinfo {author}
  {\bibfnamefont {A.}~\bibnamefont {Scholl}}, \bibinfo {author} {\bibfnamefont
  {L.~E.}\ \bibnamefont {De~Long}}, \bibinfo {author} {\bibfnamefont {J.~T.}\
  \bibnamefont {Hastings}}, \bibinfo {author} {\bibfnamefont {S.~A.}\
  \bibnamefont {Morley}},  \emph {et~al.},\ }\href@noop {} {\bibfield
  {journal} {\bibinfo  {journal} {arXiv preprint arXiv:2205.03475}\ } (\bibinfo
  {year} {2022})}\BibitemShut {NoStop}%
\bibitem [{\citenamefont {Rouxel}\ \emph {et~al.}(2022)\citenamefont {Rouxel},
  \citenamefont {R{\"o}sner}, \citenamefont {Karpov}, \citenamefont {Bacellar},
  \citenamefont {Mancini}, \citenamefont {Zinna}, \citenamefont {Kinschel},
  \citenamefont {Cannelli}, \citenamefont {Oppermann}, \citenamefont {Svetina}
  \emph {et~al.}}]{rouxel2022hard}%
  \BibitemOpen
  \bibfield  {author} {\bibinfo {author} {\bibfnamefont {J.~R.}\ \bibnamefont
  {Rouxel}}, \bibinfo {author} {\bibfnamefont {B.}~\bibnamefont {R{\"o}sner}},
  \bibinfo {author} {\bibfnamefont {D.}~\bibnamefont {Karpov}}, \bibinfo
  {author} {\bibfnamefont {C.}~\bibnamefont {Bacellar}}, \bibinfo {author}
  {\bibfnamefont {G.~F.}\ \bibnamefont {Mancini}}, \bibinfo {author}
  {\bibfnamefont {F.}~\bibnamefont {Zinna}}, \bibinfo {author} {\bibfnamefont
  {D.}~\bibnamefont {Kinschel}}, \bibinfo {author} {\bibfnamefont
  {O.}~\bibnamefont {Cannelli}}, \bibinfo {author} {\bibfnamefont
  {M.}~\bibnamefont {Oppermann}}, \bibinfo {author} {\bibfnamefont
  {C.}~\bibnamefont {Svetina}},  \emph {et~al.},\ }\href@noop {} {\bibfield
  {journal} {\bibinfo  {journal} {Nature Photonics}\ ,\ \bibinfo {pages} {1}}
  (\bibinfo {year} {2022})}\BibitemShut {NoStop}%
\bibitem [{\citenamefont {Pellegrini}\ \emph {et~al.}(2016)\citenamefont
  {Pellegrini}, \citenamefont {Marinelli},\ and\ \citenamefont
  {Reiche}}]{Pellegrinireview}%
  \BibitemOpen
  \bibfield  {author} {\bibinfo {author} {\bibfnamefont {C.}~\bibnamefont
  {Pellegrini}}, \bibinfo {author} {\bibfnamefont {A.}~\bibnamefont
  {Marinelli}}, \ and\ \bibinfo {author} {\bibfnamefont {S.}~\bibnamefont
  {Reiche}},\ }\href@noop {} {\bibfield  {journal} {\bibinfo  {journal} {Review
  of Modern Physics}\ }\textbf {\bibinfo {volume} {88}} (\bibinfo {year}
  {2016})}\BibitemShut {NoStop}%
\bibitem [{\citenamefont {Huang}\ \emph {et~al.}(2021)\citenamefont {Huang},
  \citenamefont {Deng}, \citenamefont {Liu}, \citenamefont {Wang},\ and\
  \citenamefont {Zhao}}]{Huangreview}%
  \BibitemOpen
  \bibfield  {author} {\bibinfo {author} {\bibfnamefont {N.}~\bibnamefont
  {Huang}}, \bibinfo {author} {\bibfnamefont {H.}~\bibnamefont {Deng}},
  \bibinfo {author} {\bibfnamefont {B.}~\bibnamefont {Liu}}, \bibinfo {author}
  {\bibfnamefont {D.}~\bibnamefont {Wang}}, \ and\ \bibinfo {author}
  {\bibfnamefont {Z.}~\bibnamefont {Zhao}},\ }\href@noop {} {\bibfield
  {journal} {\bibinfo  {journal} {The Innovation}\ }\textbf {\bibinfo {volume}
  {2}},\ \bibinfo {pages} {100097} (\bibinfo {year} {2021})}\BibitemShut
  {NoStop}%
\bibitem [{\citenamefont {Emma}\ \emph {et~al.}(2010)\citenamefont {Emma},
  \citenamefont {Akre}, \citenamefont {Arthur}, \citenamefont {Bionta},
  \citenamefont {Bostedt}, \citenamefont {Bozek}, \citenamefont {Brachmann},
  \citenamefont {Bucksbaum}, \citenamefont {Coffee}, \citenamefont {Decker}
  \emph {et~al.}}]{lcls}%
  \BibitemOpen
  \bibfield  {author} {\bibinfo {author} {\bibfnamefont {P.}~\bibnamefont
  {Emma}}, \bibinfo {author} {\bibfnamefont {R.}~\bibnamefont {Akre}}, \bibinfo
  {author} {\bibfnamefont {J.}~\bibnamefont {Arthur}}, \bibinfo {author}
  {\bibfnamefont {R.}~\bibnamefont {Bionta}}, \bibinfo {author} {\bibfnamefont
  {C.}~\bibnamefont {Bostedt}}, \bibinfo {author} {\bibfnamefont
  {J.}~\bibnamefont {Bozek}}, \bibinfo {author} {\bibfnamefont
  {A.}~\bibnamefont {Brachmann}}, \bibinfo {author} {\bibfnamefont
  {P.}~\bibnamefont {Bucksbaum}}, \bibinfo {author} {\bibfnamefont
  {R.}~\bibnamefont {Coffee}}, \bibinfo {author} {\bibfnamefont {F.-J.}\
  \bibnamefont {Decker}},  \emph {et~al.},\ }\href@noop {} {\bibfield
  {journal} {\bibinfo  {journal} {Nature Photonics}\ }\textbf {\bibinfo
  {volume} {4}},\ \bibinfo {pages} {641} (\bibinfo {year} {2010})}\BibitemShut
  {NoStop}%
\bibitem [{\citenamefont {Pile}(2011)}]{sacla}%
  \BibitemOpen
  \bibfield  {author} {\bibinfo {author} {\bibfnamefont {D.}~\bibnamefont
  {Pile}},\ }\href@noop {} {\bibfield  {journal} {\bibinfo  {journal} {Nature
  Photonics}\ }\textbf {\bibinfo {volume} {5}},\ \bibinfo {pages} {456}
  (\bibinfo {year} {2011})}\BibitemShut {NoStop}%
\bibitem [{\citenamefont {Kang}\ \emph {et~al.}(2017)\citenamefont {Kang},
  \citenamefont {Min}, \citenamefont {Heo}, \citenamefont {Kim}, \citenamefont
  {Yang}, \citenamefont {Kim}, \citenamefont {Nam}, \citenamefont {Baek},
  \citenamefont {Choi}, \citenamefont {Mun} \emph {et~al.}}]{pal}%
  \BibitemOpen
  \bibfield  {author} {\bibinfo {author} {\bibfnamefont {H.-S.}\ \bibnamefont
  {Kang}}, \bibinfo {author} {\bibfnamefont {C.-K.}\ \bibnamefont {Min}},
  \bibinfo {author} {\bibfnamefont {H.}~\bibnamefont {Heo}}, \bibinfo {author}
  {\bibfnamefont {C.}~\bibnamefont {Kim}}, \bibinfo {author} {\bibfnamefont
  {H.}~\bibnamefont {Yang}}, \bibinfo {author} {\bibfnamefont {G.}~\bibnamefont
  {Kim}}, \bibinfo {author} {\bibfnamefont {I.}~\bibnamefont {Nam}}, \bibinfo
  {author} {\bibfnamefont {S.~Y.}\ \bibnamefont {Baek}}, \bibinfo {author}
  {\bibfnamefont {H.-J.}\ \bibnamefont {Choi}}, \bibinfo {author}
  {\bibfnamefont {G.}~\bibnamefont {Mun}},  \emph {et~al.},\ }\href@noop {}
  {\bibfield  {journal} {\bibinfo  {journal} {Nature Photonics}\ }\textbf
  {\bibinfo {volume} {11}},\ \bibinfo {pages} {708} (\bibinfo {year}
  {2017})}\BibitemShut {NoStop}%
\bibitem [{\citenamefont {Decking}\ \emph {et~al.}(2020)\citenamefont
  {Decking}, \citenamefont {Abeghyan}, \citenamefont {Abramian}, \citenamefont
  {Abramsky}, \citenamefont {Aguirre}, \citenamefont {Albrecht}, \citenamefont
  {Alou}, \citenamefont {Altarelli}, \citenamefont {Altmann}, \citenamefont
  {Amyan} \emph {et~al.}}]{decking2020mhz}%
  \BibitemOpen
  \bibfield  {author} {\bibinfo {author} {\bibfnamefont {W.}~\bibnamefont
  {Decking}}, \bibinfo {author} {\bibfnamefont {S.}~\bibnamefont {Abeghyan}},
  \bibinfo {author} {\bibfnamefont {P.}~\bibnamefont {Abramian}}, \bibinfo
  {author} {\bibfnamefont {A.}~\bibnamefont {Abramsky}}, \bibinfo {author}
  {\bibfnamefont {A.}~\bibnamefont {Aguirre}}, \bibinfo {author} {\bibfnamefont
  {C.}~\bibnamefont {Albrecht}}, \bibinfo {author} {\bibfnamefont
  {P.}~\bibnamefont {Alou}}, \bibinfo {author} {\bibfnamefont {M.}~\bibnamefont
  {Altarelli}}, \bibinfo {author} {\bibfnamefont {P.}~\bibnamefont {Altmann}},
  \bibinfo {author} {\bibfnamefont {K.}~\bibnamefont {Amyan}},  \emph
  {et~al.},\ }\href@noop {} {\bibfield  {journal} {\bibinfo  {journal} {Nature
  Photonics}\ ,\ \bibinfo {pages} {1}} (\bibinfo {year} {2020})}\BibitemShut
  {NoStop}%
\bibitem [{\citenamefont {Prat}\ \emph {et~al.}(2020)\citenamefont {Prat},
  \citenamefont {Abela}, \citenamefont {Aiba}, \citenamefont {Alarcon},
  \citenamefont {Alex}, \citenamefont {Arbelo}, \citenamefont {Arrell},
  \citenamefont {Arsov}, \citenamefont {Bacellar}, \citenamefont {Beard} \emph
  {et~al.}}]{swissFEL}%
  \BibitemOpen
  \bibfield  {author} {\bibinfo {author} {\bibfnamefont {E.}~\bibnamefont
  {Prat}}, \bibinfo {author} {\bibfnamefont {R.}~\bibnamefont {Abela}},
  \bibinfo {author} {\bibfnamefont {M.}~\bibnamefont {Aiba}}, \bibinfo {author}
  {\bibfnamefont {A.}~\bibnamefont {Alarcon}}, \bibinfo {author} {\bibfnamefont
  {J.}~\bibnamefont {Alex}}, \bibinfo {author} {\bibfnamefont {Y.}~\bibnamefont
  {Arbelo}}, \bibinfo {author} {\bibfnamefont {C.}~\bibnamefont {Arrell}},
  \bibinfo {author} {\bibfnamefont {V.}~\bibnamefont {Arsov}}, \bibinfo
  {author} {\bibfnamefont {C.}~\bibnamefont {Bacellar}}, \bibinfo {author}
  {\bibfnamefont {C.}~\bibnamefont {Beard}},  \emph {et~al.},\ }\href@noop {}
  {\bibfield  {journal} {\bibinfo  {journal} {Nature Photonics}\ }\textbf
  {\bibinfo {volume} {14}},\ \bibinfo {pages} {748} (\bibinfo {year}
  {2020})}\BibitemShut {NoStop}%
\bibitem [{\citenamefont {Kondratenko}\ and\ \citenamefont
  {Saldin}(1980)}]{sase}%
  \BibitemOpen
  \bibfield  {author} {\bibinfo {author} {\bibfnamefont {A.}~\bibnamefont
  {Kondratenko}}\ and\ \bibinfo {author} {\bibfnamefont {E.}~\bibnamefont
  {Saldin}},\ }\href@noop {} {\bibfield  {journal} {\bibinfo  {journal} {Part.
  Accel.}\ }\textbf {\bibinfo {volume} {10}},\ \bibinfo {pages} {207} (\bibinfo
  {year} {1980})}\BibitemShut {NoStop}%
\bibitem [{\citenamefont {Feldhaus}\ \emph {et~al.}(1997)\citenamefont
  {Feldhaus}, \citenamefont {Saldin}, \citenamefont {Schneider}, \citenamefont
  {Schneidmiller},\ and\ \citenamefont {Yurkov}}]{feldhaus1997possible}%
  \BibitemOpen
  \bibfield  {author} {\bibinfo {author} {\bibfnamefont {J.}~\bibnamefont
  {Feldhaus}}, \bibinfo {author} {\bibfnamefont {E.}~\bibnamefont {Saldin}},
  \bibinfo {author} {\bibfnamefont {J.}~\bibnamefont {Schneider}}, \bibinfo
  {author} {\bibfnamefont {E.}~\bibnamefont {Schneidmiller}}, \ and\ \bibinfo
  {author} {\bibfnamefont {M.}~\bibnamefont {Yurkov}},\ }\href@noop {}
  {\bibfield  {journal} {\bibinfo  {journal} {Optics Communications}\ }\textbf
  {\bibinfo {volume} {140}},\ \bibinfo {pages} {341} (\bibinfo {year}
  {1997})}\BibitemShut {NoStop}%
\bibitem [{\citenamefont {Geloni}\ \emph {et~al.}(2011)\citenamefont {Geloni},
  \citenamefont {Kocharyan},\ and\ \citenamefont {Saldin}}]{geloni2011novel}%
  \BibitemOpen
  \bibfield  {author} {\bibinfo {author} {\bibfnamefont {G.}~\bibnamefont
  {Geloni}}, \bibinfo {author} {\bibfnamefont {V.}~\bibnamefont {Kocharyan}}, \
  and\ \bibinfo {author} {\bibfnamefont {E.}~\bibnamefont {Saldin}},\
  }\href@noop {} {\bibfield  {journal} {\bibinfo  {journal} {Journal of Modern
  Optics}\ }\textbf {\bibinfo {volume} {58}},\ \bibinfo {pages} {1391}
  (\bibinfo {year} {2011})}\BibitemShut {NoStop}%
\bibitem [{\citenamefont {Colson}(1981)}]{colson1981nonlinear}%
  \BibitemOpen
  \bibfield  {author} {\bibinfo {author} {\bibfnamefont {W.}~\bibnamefont
  {Colson}},\ }\href@noop {} {\bibfield  {journal} {\bibinfo  {journal} {IEEE
  Journal of Quantum Electronics}\ }\textbf {\bibinfo {volume} {17}},\ \bibinfo
  {pages} {1417} (\bibinfo {year} {1981})}\BibitemShut {NoStop}%
\bibitem [{\citenamefont {Geloni}\ \emph {et~al.}(2007)\citenamefont {Geloni},
  \citenamefont {Saldin}, \citenamefont {Schneidmiller},\ and\ \citenamefont
  {Yurkov}}]{geloni2007theory}%
  \BibitemOpen
  \bibfield  {author} {\bibinfo {author} {\bibfnamefont {G.}~\bibnamefont
  {Geloni}}, \bibinfo {author} {\bibfnamefont {E.}~\bibnamefont {Saldin}},
  \bibinfo {author} {\bibfnamefont {E.}~\bibnamefont {Schneidmiller}}, \ and\
  \bibinfo {author} {\bibfnamefont {M.}~\bibnamefont {Yurkov}},\ }\href@noop {}
  {\bibfield  {journal} {\bibinfo  {journal} {Nuclear Instruments and Methods
  in Physics Research Section A: Accelerators, Spectrometers, Detectors and
  Associated Equipment}\ }\textbf {\bibinfo {volume} {581}},\ \bibinfo {pages}
  {856} (\bibinfo {year} {2007})}\BibitemShut {NoStop}%
\bibitem [{\citenamefont {Hemsing}(2020)}]{hemsing2020coherent}%
  \BibitemOpen
  \bibfield  {author} {\bibinfo {author} {\bibfnamefont {E.}~\bibnamefont
  {Hemsing}},\ }\href@noop {} {\bibfield  {journal} {\bibinfo  {journal}
  {Physical Review Accelerators and Beams}\ }\textbf {\bibinfo {volume} {23}},\
  \bibinfo {pages} {020703} (\bibinfo {year} {2020})}\BibitemShut {NoStop}%
\bibitem [{\citenamefont {Bahrdt}\ \emph {et~al.}(2013)\citenamefont {Bahrdt},
  \citenamefont {Holldack}, \citenamefont {Kuske}, \citenamefont {M{\"u}ller},
  \citenamefont {Scheer},\ and\ \citenamefont {Schmid}}]{bahrdt2013first}%
  \BibitemOpen
  \bibfield  {author} {\bibinfo {author} {\bibfnamefont {J.}~\bibnamefont
  {Bahrdt}}, \bibinfo {author} {\bibfnamefont {K.}~\bibnamefont {Holldack}},
  \bibinfo {author} {\bibfnamefont {P.}~\bibnamefont {Kuske}}, \bibinfo
  {author} {\bibfnamefont {R.}~\bibnamefont {M{\"u}ller}}, \bibinfo {author}
  {\bibfnamefont {M.}~\bibnamefont {Scheer}}, \ and\ \bibinfo {author}
  {\bibfnamefont {P.}~\bibnamefont {Schmid}},\ }\href@noop {} {\bibfield
  {journal} {\bibinfo  {journal} {Physical Review Letters}\ }\textbf {\bibinfo
  {volume} {111}},\ \bibinfo {pages} {034801} (\bibinfo {year}
  {2013})}\BibitemShut {NoStop}%
\bibitem [{\citenamefont {Hemsing}\ \emph {et~al.}(2014)\citenamefont
  {Hemsing}, \citenamefont {Dunning}, \citenamefont {Hast}, \citenamefont
  {Raubenheimer},\ and\ \citenamefont {Xiang}}]{hemsing2014first}%
  \BibitemOpen
  \bibfield  {author} {\bibinfo {author} {\bibfnamefont {E.}~\bibnamefont
  {Hemsing}}, \bibinfo {author} {\bibfnamefont {M.}~\bibnamefont {Dunning}},
  \bibinfo {author} {\bibfnamefont {C.}~\bibnamefont {Hast}}, \bibinfo {author}
  {\bibfnamefont {T.}~\bibnamefont {Raubenheimer}}, \ and\ \bibinfo {author}
  {\bibfnamefont {D.}~\bibnamefont {Xiang}},\ }\href@noop {} {\bibfield
  {journal} {\bibinfo  {journal} {Physical Review Letters}\ }\textbf {\bibinfo
  {volume} {113}},\ \bibinfo {pages} {134803} (\bibinfo {year}
  {2014})}\BibitemShut {NoStop}%
\bibitem [{\citenamefont {Ribi{\v{c}}}\ \emph {et~al.}(2017)\citenamefont
  {Ribi{\v{c}}}, \citenamefont {R{\"o}sner}, \citenamefont {Gauthier},
  \citenamefont {Allaria}, \citenamefont {D{\"o}ring}, \citenamefont {Foglia},
  \citenamefont {Giannessi}, \citenamefont {Mahne}, \citenamefont {Manfredda},
  \citenamefont {Masciovecchio} \emph {et~al.}}]{ribivc2017extreme}%
  \BibitemOpen
  \bibfield  {author} {\bibinfo {author} {\bibfnamefont {P.~R.}\ \bibnamefont
  {Ribi{\v{c}}}}, \bibinfo {author} {\bibfnamefont {B.}~\bibnamefont
  {R{\"o}sner}}, \bibinfo {author} {\bibfnamefont {D.}~\bibnamefont
  {Gauthier}}, \bibinfo {author} {\bibfnamefont {E.}~\bibnamefont {Allaria}},
  \bibinfo {author} {\bibfnamefont {F.}~\bibnamefont {D{\"o}ring}}, \bibinfo
  {author} {\bibfnamefont {L.}~\bibnamefont {Foglia}}, \bibinfo {author}
  {\bibfnamefont {L.}~\bibnamefont {Giannessi}}, \bibinfo {author}
  {\bibfnamefont {N.}~\bibnamefont {Mahne}}, \bibinfo {author} {\bibfnamefont
  {M.}~\bibnamefont {Manfredda}}, \bibinfo {author} {\bibfnamefont
  {C.}~\bibnamefont {Masciovecchio}},  \emph {et~al.},\ }\href@noop {}
  {\bibfield  {journal} {\bibinfo  {journal} {Physical Review X}\ }\textbf
  {\bibinfo {volume} {7}},\ \bibinfo {pages} {031036} (\bibinfo {year}
  {2017})}\BibitemShut {NoStop}%
\bibitem [{\citenamefont {Hemsing}\ \emph {et~al.}(2011)\citenamefont
  {Hemsing}, \citenamefont {Marinelli},\ and\ \citenamefont
  {Rosenzweig}}]{hemsing2011generating}%
  \BibitemOpen
  \bibfield  {author} {\bibinfo {author} {\bibfnamefont {E.}~\bibnamefont
  {Hemsing}}, \bibinfo {author} {\bibfnamefont {A.}~\bibnamefont {Marinelli}},
  \ and\ \bibinfo {author} {\bibfnamefont {J.}~\bibnamefont {Rosenzweig}},\
  }\href@noop {} {\bibfield  {journal} {\bibinfo  {journal} {Physical Review
  Letters}\ }\textbf {\bibinfo {volume} {106}},\ \bibinfo {pages} {164803}
  (\bibinfo {year} {2011})}\BibitemShut {NoStop}%
\bibitem [{\citenamefont {Hemsing}\ \emph {et~al.}(2013)\citenamefont
  {Hemsing}, \citenamefont {Knyazik}, \citenamefont {Dunning}, \citenamefont
  {Xiang}, \citenamefont {Marinelli}, \citenamefont {Hast},\ and\ \citenamefont
  {Rosenzweig}}]{hemsing2013coherent}%
  \BibitemOpen
  \bibfield  {author} {\bibinfo {author} {\bibfnamefont {E.}~\bibnamefont
  {Hemsing}}, \bibinfo {author} {\bibfnamefont {A.}~\bibnamefont {Knyazik}},
  \bibinfo {author} {\bibfnamefont {M.}~\bibnamefont {Dunning}}, \bibinfo
  {author} {\bibfnamefont {D.}~\bibnamefont {Xiang}}, \bibinfo {author}
  {\bibfnamefont {A.}~\bibnamefont {Marinelli}}, \bibinfo {author}
  {\bibfnamefont {C.}~\bibnamefont {Hast}}, \ and\ \bibinfo {author}
  {\bibfnamefont {J.~B.}\ \bibnamefont {Rosenzweig}},\ }\href@noop {}
  {\bibfield  {journal} {\bibinfo  {journal} {Nature Physics}\ }\textbf
  {\bibinfo {volume} {9}},\ \bibinfo {pages} {549} (\bibinfo {year}
  {2013})}\BibitemShut {NoStop}%
\bibitem [{\citenamefont {Hemsing}\ and\ \citenamefont
  {Marinelli}(2012)}]{hemsing2012echo}%
  \BibitemOpen
  \bibfield  {author} {\bibinfo {author} {\bibfnamefont {E.}~\bibnamefont
  {Hemsing}}\ and\ \bibinfo {author} {\bibfnamefont {A.}~\bibnamefont
  {Marinelli}},\ }\href@noop {} {\bibfield  {journal} {\bibinfo  {journal}
  {Physical Review Letters}\ }\textbf {\bibinfo {volume} {109}},\ \bibinfo
  {pages} {224801} (\bibinfo {year} {2012})}\BibitemShut {NoStop}%
\bibitem [{\citenamefont {Ribi{\v{c}}}\ \emph {et~al.}(2014)\citenamefont
  {Ribi{\v{c}}}, \citenamefont {Gauthier},\ and\ \citenamefont
  {De~Ninno}}]{ribivc2014generation}%
  \BibitemOpen
  \bibfield  {author} {\bibinfo {author} {\bibfnamefont {P.~R.}\ \bibnamefont
  {Ribi{\v{c}}}}, \bibinfo {author} {\bibfnamefont {D.}~\bibnamefont
  {Gauthier}}, \ and\ \bibinfo {author} {\bibfnamefont {G.}~\bibnamefont
  {De~Ninno}},\ }\href@noop {} {\bibfield  {journal} {\bibinfo  {journal}
  {Physical Review Letters}\ }\textbf {\bibinfo {volume} {112}},\ \bibinfo
  {pages} {203602} (\bibinfo {year} {2014})}\BibitemShut {NoStop}%
\bibitem [{\citenamefont {Huang}\ and\ \citenamefont
  {Deng}(2021)}]{huang2021generating}%
  \BibitemOpen
  \bibfield  {author} {\bibinfo {author} {\bibfnamefont {N.}~\bibnamefont
  {Huang}}\ and\ \bibinfo {author} {\bibfnamefont {H.}~\bibnamefont {Deng}},\
  }\href@noop {} {\bibfield  {journal} {\bibinfo  {journal} {Optica}\ }\textbf
  {\bibinfo {volume} {8}},\ \bibinfo {pages} {1020} (\bibinfo {year}
  {2021})}\BibitemShut {NoStop}%
\bibitem [{\citenamefont {Peele}\ \emph {et~al.}(2002)\citenamefont {Peele},
  \citenamefont {McMahon}, \citenamefont {Paterson}, \citenamefont {Tran},
  \citenamefont {Mancuso}, \citenamefont {Nugent}, \citenamefont {Hayes},
  \citenamefont {Harvey}, \citenamefont {Lai},\ and\ \citenamefont
  {McNulty}}]{peele2002observation}%
  \BibitemOpen
  \bibfield  {author} {\bibinfo {author} {\bibfnamefont {A.~G.}\ \bibnamefont
  {Peele}}, \bibinfo {author} {\bibfnamefont {P.~J.}\ \bibnamefont {McMahon}},
  \bibinfo {author} {\bibfnamefont {D.}~\bibnamefont {Paterson}}, \bibinfo
  {author} {\bibfnamefont {C.~Q.}\ \bibnamefont {Tran}}, \bibinfo {author}
  {\bibfnamefont {A.~P.}\ \bibnamefont {Mancuso}}, \bibinfo {author}
  {\bibfnamefont {K.~A.}\ \bibnamefont {Nugent}}, \bibinfo {author}
  {\bibfnamefont {J.~P.}\ \bibnamefont {Hayes}}, \bibinfo {author}
  {\bibfnamefont {E.}~\bibnamefont {Harvey}}, \bibinfo {author} {\bibfnamefont
  {B.}~\bibnamefont {Lai}}, \ and\ \bibinfo {author} {\bibfnamefont
  {I.}~\bibnamefont {McNulty}},\ }\href@noop {} {\bibfield  {journal} {\bibinfo
   {journal} {Optics Letters}\ }\textbf {\bibinfo {volume} {27}},\ \bibinfo
  {pages} {1752} (\bibinfo {year} {2002})}\BibitemShut {NoStop}%
\bibitem [{\citenamefont {Seiboth}\ \emph {et~al.}(2019)\citenamefont
  {Seiboth}, \citenamefont {Kahnt}, \citenamefont {Lyubomirskiy}, \citenamefont
  {Seyrich}, \citenamefont {Wittwer}, \citenamefont {Ullsperger}, \citenamefont
  {Nolte}, \citenamefont {Batey}, \citenamefont {Rau},\ and\ \citenamefont
  {Schroer}}]{seiboth2019refractive}%
  \BibitemOpen
  \bibfield  {author} {\bibinfo {author} {\bibfnamefont {F.}~\bibnamefont
  {Seiboth}}, \bibinfo {author} {\bibfnamefont {M.}~\bibnamefont {Kahnt}},
  \bibinfo {author} {\bibfnamefont {M.}~\bibnamefont {Lyubomirskiy}}, \bibinfo
  {author} {\bibfnamefont {M.}~\bibnamefont {Seyrich}}, \bibinfo {author}
  {\bibfnamefont {F.}~\bibnamefont {Wittwer}}, \bibinfo {author} {\bibfnamefont
  {T.}~\bibnamefont {Ullsperger}}, \bibinfo {author} {\bibfnamefont
  {S.}~\bibnamefont {Nolte}}, \bibinfo {author} {\bibfnamefont
  {D.}~\bibnamefont {Batey}}, \bibinfo {author} {\bibfnamefont
  {C.}~\bibnamefont {Rau}}, \ and\ \bibinfo {author} {\bibfnamefont {C.~G.}\
  \bibnamefont {Schroer}},\ }\href@noop {} {\bibfield  {journal} {\bibinfo
  {journal} {Optics Letters}\ }\textbf {\bibinfo {volume} {44}},\ \bibinfo
  {pages} {4622} (\bibinfo {year} {2019})}\BibitemShut {NoStop}%
\bibitem [{\citenamefont {Sakdinawat}\ and\ \citenamefont
  {Liu}(2007)}]{sakdinawat2007soft}%
  \BibitemOpen
  \bibfield  {author} {\bibinfo {author} {\bibfnamefont {A.}~\bibnamefont
  {Sakdinawat}}\ and\ \bibinfo {author} {\bibfnamefont {Y.}~\bibnamefont
  {Liu}},\ }\href@noop {} {\bibfield  {journal} {\bibinfo  {journal} {Optics
  Letters}\ }\textbf {\bibinfo {volume} {32}},\ \bibinfo {pages} {2635}
  (\bibinfo {year} {2007})}\BibitemShut {NoStop}%
\bibitem [{\citenamefont {Vila-Comamala}\ \emph {et~al.}(2014)\citenamefont
  {Vila-Comamala}, \citenamefont {Sakdinawat},\ and\ \citenamefont
  {Guizar-Sicairos}}]{vila2014characterization}%
  \BibitemOpen
  \bibfield  {author} {\bibinfo {author} {\bibfnamefont {J.}~\bibnamefont
  {Vila-Comamala}}, \bibinfo {author} {\bibfnamefont {A.}~\bibnamefont
  {Sakdinawat}}, \ and\ \bibinfo {author} {\bibfnamefont {M.}~\bibnamefont
  {Guizar-Sicairos}},\ }\href@noop {} {\bibfield  {journal} {\bibinfo
  {journal} {Optics Letters}\ }\textbf {\bibinfo {volume} {39}},\ \bibinfo
  {pages} {5281} (\bibinfo {year} {2014})}\BibitemShut {NoStop}%
\bibitem [{\citenamefont {Lee}\ \emph {et~al.}(2019)\citenamefont {Lee},
  \citenamefont {Alexander}, \citenamefont {Kevan}, \citenamefont {Roy},\ and\
  \citenamefont {McMorran}}]{lee2019laguerre}%
  \BibitemOpen
  \bibfield  {author} {\bibinfo {author} {\bibfnamefont {J.~T.}\ \bibnamefont
  {Lee}}, \bibinfo {author} {\bibfnamefont {S.}~\bibnamefont {Alexander}},
  \bibinfo {author} {\bibfnamefont {S.}~\bibnamefont {Kevan}}, \bibinfo
  {author} {\bibfnamefont {S.}~\bibnamefont {Roy}}, \ and\ \bibinfo {author}
  {\bibfnamefont {B.}~\bibnamefont {McMorran}},\ }\href@noop {} {\bibfield
  {journal} {\bibinfo  {journal} {Nature Photonics}\ }\textbf {\bibinfo
  {volume} {13}},\ \bibinfo {pages} {205} (\bibinfo {year} {2019})}\BibitemShut
  {NoStop}%
\bibitem [{\citenamefont {Hemsing}\ \emph
  {et~al.}(2008{\natexlab{a}})\citenamefont {Hemsing}, \citenamefont
  {Marinelli}, \citenamefont {Reiche},\ and\ \citenamefont
  {Rosenzweig}}]{hemsing2008longitudinal}%
  \BibitemOpen
  \bibfield  {author} {\bibinfo {author} {\bibfnamefont {E.}~\bibnamefont
  {Hemsing}}, \bibinfo {author} {\bibfnamefont {A.}~\bibnamefont {Marinelli}},
  \bibinfo {author} {\bibfnamefont {S.}~\bibnamefont {Reiche}}, \ and\ \bibinfo
  {author} {\bibfnamefont {J.}~\bibnamefont {Rosenzweig}},\ }\href@noop {}
  {\bibfield  {journal} {\bibinfo  {journal} {Physical Review Special
  Topics-Accelerators and Beams}\ }\textbf {\bibinfo {volume} {11}},\ \bibinfo
  {pages} {070704} (\bibinfo {year} {2008}{\natexlab{a}})}\BibitemShut
  {NoStop}%
\bibitem [{\citenamefont {Lutman}\ \emph {et~al.}(2016)\citenamefont {Lutman},
  \citenamefont {Maxwell}, \citenamefont {MacArthur}, \citenamefont {Guetg},
  \citenamefont {Berrah}, \citenamefont {Coffee}, \citenamefont {Ding},
  \citenamefont {Huang}, \citenamefont {Marinelli}, \citenamefont {Moeller}
  \emph {et~al.}}]{lutman2016fresh}%
  \BibitemOpen
  \bibfield  {author} {\bibinfo {author} {\bibfnamefont {A.~A.}\ \bibnamefont
  {Lutman}}, \bibinfo {author} {\bibfnamefont {T.~J.}\ \bibnamefont {Maxwell}},
  \bibinfo {author} {\bibfnamefont {J.~P.}\ \bibnamefont {MacArthur}}, \bibinfo
  {author} {\bibfnamefont {M.~W.}\ \bibnamefont {Guetg}}, \bibinfo {author}
  {\bibfnamefont {N.}~\bibnamefont {Berrah}}, \bibinfo {author} {\bibfnamefont
  {R.~N.}\ \bibnamefont {Coffee}}, \bibinfo {author} {\bibfnamefont
  {Y.}~\bibnamefont {Ding}}, \bibinfo {author} {\bibfnamefont {Z.}~\bibnamefont
  {Huang}}, \bibinfo {author} {\bibfnamefont {A.}~\bibnamefont {Marinelli}},
  \bibinfo {author} {\bibfnamefont {S.}~\bibnamefont {Moeller}},  \emph
  {et~al.},\ }\href@noop {} {\bibfield  {journal} {\bibinfo  {journal} {Nature
  Photonics}\ }\textbf {\bibinfo {volume} {10}},\ \bibinfo {pages} {745}
  (\bibinfo {year} {2016})}\BibitemShut {NoStop}%
\bibitem [{\citenamefont {Marinelli}\ \emph {et~al.}(2015)\citenamefont
  {Marinelli}, \citenamefont {Ratner}, \citenamefont {Lutman}, \citenamefont
  {Turner}, \citenamefont {Welch}, \citenamefont {Decker}, \citenamefont
  {Loos}, \citenamefont {Behrens}, \citenamefont {Gilevich}, \citenamefont
  {Miahnahri} \emph {et~al.}}]{marinelli2015high}%
  \BibitemOpen
  \bibfield  {author} {\bibinfo {author} {\bibfnamefont {A.}~\bibnamefont
  {Marinelli}}, \bibinfo {author} {\bibfnamefont {D.}~\bibnamefont {Ratner}},
  \bibinfo {author} {\bibfnamefont {A.}~\bibnamefont {Lutman}}, \bibinfo
  {author} {\bibfnamefont {J.}~\bibnamefont {Turner}}, \bibinfo {author}
  {\bibfnamefont {J.}~\bibnamefont {Welch}}, \bibinfo {author} {\bibfnamefont
  {F.-J.}\ \bibnamefont {Decker}}, \bibinfo {author} {\bibfnamefont
  {H.}~\bibnamefont {Loos}}, \bibinfo {author} {\bibfnamefont {C.}~\bibnamefont
  {Behrens}}, \bibinfo {author} {\bibfnamefont {S.}~\bibnamefont {Gilevich}},
  \bibinfo {author} {\bibfnamefont {A.}~\bibnamefont {Miahnahri}},  \emph
  {et~al.},\ }\href@noop {} {\bibfield  {journal} {\bibinfo  {journal} {Nature
  Communications}\ }\textbf {\bibinfo {volume} {6}},\ \bibinfo {pages} {1}
  (\bibinfo {year} {2015})}\BibitemShut {NoStop}%
\bibitem [{\citenamefont {Hemsing}\ \emph
  {et~al.}(2008{\natexlab{b}})\citenamefont {Hemsing}, \citenamefont {Gover},\
  and\ \citenamefont {Rosenzweig}}]{hemsing2008virtual}%
  \BibitemOpen
  \bibfield  {author} {\bibinfo {author} {\bibfnamefont {E.}~\bibnamefont
  {Hemsing}}, \bibinfo {author} {\bibfnamefont {A.}~\bibnamefont {Gover}}, \
  and\ \bibinfo {author} {\bibfnamefont {J.}~\bibnamefont {Rosenzweig}},\
  }\href@noop {} {\bibfield  {journal} {\bibinfo  {journal} {Physical Review
  A}\ }\textbf {\bibinfo {volume} {77}},\ \bibinfo {pages} {063831} (\bibinfo
  {year} {2008}{\natexlab{b}})}\BibitemShut {NoStop}%
\bibitem [{\citenamefont {Saldin}\ \emph {et~al.}(2000)\citenamefont {Saldin},
  \citenamefont {Schneidmiller},\ and\ \citenamefont
  {Yurkov}}]{saldin2000diffraction}%
  \BibitemOpen
  \bibfield  {author} {\bibinfo {author} {\bibfnamefont {E.}~\bibnamefont
  {Saldin}}, \bibinfo {author} {\bibfnamefont {E.}~\bibnamefont
  {Schneidmiller}}, \ and\ \bibinfo {author} {\bibfnamefont {M.}~\bibnamefont
  {Yurkov}},\ }\href@noop {} {\bibfield  {journal} {\bibinfo  {journal} {Optics
  Communications}\ }\textbf {\bibinfo {volume} {186}},\ \bibinfo {pages} {185}
  (\bibinfo {year} {2000})}\BibitemShut {NoStop}%
\bibitem [{\citenamefont {Reiche}(1999)}]{reiche1999genesis}%
  \BibitemOpen
  \bibfield  {author} {\bibinfo {author} {\bibfnamefont {S.}~\bibnamefont
  {Reiche}},\ }\href@noop {} {\bibfield  {journal} {\bibinfo  {journal}
  {Nuclear Instruments and Methods in Physics Research Section A: Accelerators,
  Spectrometers, Detectors and Associated Equipment}\ }\textbf {\bibinfo
  {volume} {429}},\ \bibinfo {pages} {243} (\bibinfo {year}
  {1999})}\BibitemShut {NoStop}%
\end{thebibliography}%

\end{document}